\documentclass[final,1p,times]{elsarticle}
\usepackage{amsmath,amssymb}
\usepackage{upgreek}
\usepackage[noend]{algpseudocode}
\usepackage{algorithm,algorithmicx}
\usepackage{pgf,tikz}
\usetikzlibrary{shapes,arrows,automata}
\usepackage{textcomp}
\usepackage{hyperref}
\usepackage{array}
\usepackage[braket,qm]{qcircuit}
\usepackage{dsfont}
\usepackage{lineno}

\newtheorem{thm2}{Theorem}[section]

\newtheorem{definition}[thm2]{Definition}
\newtheorem{lemma}[thm2]{Lemma}
\newtheorem{theorem}[thm2]{Theorem}
\newtheorem{remark}[thm2]{Remark}
\newtheorem{example}[thm2]{Example}
\newproof{proof}{Proof}

\newcommand{\QMC}{\mathfrak{C}}
\newcommand{\h}{\mathcal{H}}
\newcommand{\LLL}{\mathcal{L}}
\newcommand{\DDD}{\mathcal{D}}
\newcommand{\SSS}{\mathcal{S}}
\newcommand{\EEE}{\mathcal{E}}
\newcommand{\EE}{\mathbf{E}}
\newcommand{\FFF}{\mathcal{F}}
\newcommand{\FF}{\mathbf{F}}
\newcommand{\PPP}{\mathcal{P}}
\newcommand{\PP}{\mathbf{P}}
\newcommand{\III}{\mathcal{I}}
\newcommand{\M}{\mathds{M}}
\newcommand{\ind}{\mathbf{I}}
\newcommand{\fid}{\mathrm{Fid}}
\newcommand{\mfid}{\underline{\mathrm{Fid}}}
\newcommand{\QQ}{\mathfrak{F}}
\newcommand{\vm}{\mathrm{V2L}}
\newcommand{\mv}{\mathrm{L2V}}
\newcommand{\sm}{\mathrm{S2M}}
\newcommand{\cq}{\mathrm{cq}}

\newcommand{\tr}{\mathrm{tr}}
\newcommand{\spn}{\mathrm{span}}
\newcommand{\supp}{\mathrm{supp}}

\newcommand{\BigO}{\mathcal{O}}
\newcommand{\ntl}{\mathrm{U}\,}
\newcommand{\nxt}{\mathrm{X}\,}
\newcommand{\Q}{\mathrm{Q}}
\newcommand{\x}{\mathbf{x}}
\newcommand{\z}{\mathbf{z}}
\newcommand{\true}{\mathtt{true}}
\newcommand{\false}{\mathtt{false}}
\newcommand{\ok}{\mathtt{ok}}
\newcommand{\error}{\mathtt{error}}

\algnewcommand\algorithmiccom{\textbf{Complexity:}}
\algnewcommand\Com{\item[\algorithmiccom]}

\journal{}

\begin{document}

\begin{frontmatter}



\title{An Algebraic Method to Fidelity-based
Model Checking over Quantum Markov Chains}


\author{Ming Xu}
\ead{mxu@cs.ecnu.edu.cn}
\author{Jianling Fu}
\ead{51194506011@stu.ecnu.edu.cn}
\author{Jingyi Mei}
\ead{51205902017@stu.ecnu.edu.cn}
\author{Yuxin Deng}
\ead{yxdeng@sei.ecnu.edu.cn}

\address{Shanghai Key Lab of Trustworthy Computing, East China Normal University,
Shanghai, China}

\begin{abstract}
Fidelity is one of the most widely used quantities in quantum information
that measure the distance of quantum states through a noisy channel.
In this paper,
we introduce a quantum analogy of computation tree logic (CTL) called QCTL,
which concerns fidelity instead of probability in probabilistic CTL,
over quantum Markov chains (QMCs).
Noisy channels are modelled by super-operators,
which are specified by QCTL formulas;
the initial quantum states are modelled by density operators,
which are left parametric in the given QMC.
The problem is to compute the minimum fidelity
over all initial states for conservation.
We achieve it by a reduction to quantifier elimination
in the existential theory of the reals.
The method is absolutely exact,
so that QCTL formulas are proven to be decidable in exponential time.
Finally, we implement the proposed method
and demonstrate its effectiveness via a quantum IPv4 protocol.
\end{abstract}

\begin{keyword}
Model Checking
\sep Formal Logic
\sep Quantum Computation
\sep Computer Algebra
\end{keyword}

\end{frontmatter}

\section{Introduction}
Markov chains (MCs) have attracted a lot of attention
in the field of formal verification~\cite{CGP99,BaK08}.
In 1989, Hansson and Jonsson introduced probabilistic computation tree logic (PCTL)
to specify quantitative properties over MCs,
and presented an algorithm to
check whether a property $\phi$ holds over a MC $\mathfrak{M}$~\cite{HaJ89}.
The complexity is polynomial time w.r.t. the size of both $\phi$ and $\mathfrak{M}$.
Later, more efficient approximation algorithms
were presented and implemented in various model checkers,
such as \textsc{PRISM}~\cite{KNP11},
\textsc{iscasMC}~\cite{HLS+14}
and \textsc{Storm}~\cite{DJK+17},
to solve numerous practical problems.
Such model checkers provide a Boolean answer to the decision problem:
either $\mathfrak{M}$ satisfies $\phi$ or not.
In case of a negative answer,
a counter-example can further be provided~\cite{HKD09} to locate the potential bug.
Thereby, the model checking technology has achieved great success
in both academic and industrial communities.

Quantum hardware has been rapidly developed in the last decades,
particularly in very recent years.
For example, in October 2019 Google officially announced that
its 53-qubit Sycamore processor took about 200 seconds
to sample one instance of a quantum circuit
that would have taken the world's most powerful supercomputer 10,000 years~\cite{AAB+19}.
People tend to believe that special-purpose quantum computers with more than 100 qubits
will be available in nearly 5 years.
In the meantime,
quantum software will be crucial in harnessing the power of quantum computers,
such as the BB84 protocol for quantum key distribution~\cite{BeB84},
Shor's algorithm for integer factorization~\cite{Sho94},
Grover's algorithm for unstructured search~\cite{Gro96},
and the HHL algorithm for solving linear equations~\cite{HHL09}.
To ensure the reliability of quantum software,
verification technologies are urgent to be developed for quantum systems and protocols.
Due to the features in quantum mechanics,
three major challenges in verification are:
\begin{enumerate}
	\item the state space is a continuum, 
	\item quantum information cannot be cloned~\cite{WoZ82}, and
	\item measurement destroys quantum information.
\end{enumerate}

To tackle them effectively, researchers had to impose restrictions on the quantum model.
Gay~\textit{et~al.}~\cite{GNP06,GNP08}
restricted the quantum operations to the Clifford group gates
(including Hadamard, CNOT and phase gates),
restricted the state space to a set of finitely describable states
called stabilizers
that is closed under those Clifford group gates,
and applied \textsc{PRISM} to check the quantum protocols---superdense coding,
quantum teleportation,
and quantum error correction.
Whereas, Feng~\textit{et~al.} proposed
the model of super-operator weighted Markov chain~\cite{FYY13},
which gave rise to an alternative way to finitely describable states.
The model was shown to be able to describe some hybrid systems~\cite{LiF15}.
Under the model, the authors considered the reachability probability~\cite{YFY+13},
the repeated reachability probability~\cite{FHT+17},
and the model checking of linear time properties~\cite{LiF15}
and a quantum analogy of computation tree logic (QCTL)~\cite{FYY13}.
A key step in their work is
decomposing the state space (known as a Hilbert space)
into a direct-sum of some bottom strongly connected component (BSCC) subspaces
plus a maximal transient subspace
w.r.t. a given super-operator.
After decomposition,
all the aforementioned problems were shown
to be computable/decidable in polynomial time.

The above works studied only the probability measure of some properties,
which is characterized by the trace of the final partial density operator.
Specifically, suppose a quantum system is in the state $\rho$
and some quantum channel $\EEE$ occurs,
changing the quantum system to the state $\EEE(\rho)$.
The probability measure concerns merely $\tr(\rho)$ and $\tr(\EEE(\rho))$,
which are abstractions on the whole $\rho$ and $\EEE(\rho)$.
For instance, the quantum states $\rho=\op{0}{0}$
and $\EEE(\rho)=\op{1}{1}$ (where $\EEE$ is the bit flip)
have the same probability/trace $1$,
but they are entirely different.
In other words, we fail to detect the effect of the bit flip channel.
The reason is that, in the abstraction from $\rho$ to $\tr(\rho)$,
a lot of information concerning the quantum state is lost.
In the occasions of reasoning about noisy channels,
this is far from being satisfactory.
The current work
proposes to use fidelity in place of probability measure
to specify the properties of quantum Markov chains.
Fidelity is a basic concept in quantum information that
prescribes the quantification of the similarity degree of two quantum states.
As a measure for the distance between the quantum states $\rho$ and $\EEE(\rho)$,
the fidelity, ranging in $[0,1]$,
characterizes precisely
how well a quantum channel $\EEE$ could preserve
the information of the quantum system.
Qualitatively, the fidelity is nonnegative, vanishes
if and only if $\rho$ and $\EEE(\rho)$ have support on mutually orthogonal subspaces,
and attains its maximum value $1$ if and only if the two states are identical.
It decreases as two states become more distinguishable,
where the distinguishability reflects the effect of a quantum channel.
For instance, the fidelity between $\op{0}{0}$ and $\op{1}{1}$
reaches the minimum $0$ as expected.
Hence the probability measure does not suffice to recognize general quantum states,
but the fidelity does!

In this paper, we consider the fidelity-based property
over (super-operator weighted) quantum Markov chains (QMCs).
This property is specified by another quantum analogy of computation tree logic (QCTL),
including a novel kind of fidelity-quantifier formula
instead of the trace-quantifier formula in~\cite{FYY13}.
Since the state formulas and the path formulas in QCTL are mutually inductive,
we perform the model checking in three steps:
i)~decide the basic state formulas,
ii)~synthesize the super-operators of path formulas,
and iii)~decide the fidelity-quantifier formulas.
The last step plays a central role in the model checking,
and depends on the second step.
To solve it, we first remove the BSCC subspaces
that cover all fixed-points of a super-operator in consideration.
By Brouwer's fixed-point theorem,
the direct-sum of all these BSCC subspaces are easily obtained.
Then we explicitly express the super-operators using matrix representation.
Finally the fidelity-quantifier formula is decided
by a reduction to quantifier elimination in the existential theory of the reals.
The complexity is shown to
i) be exponential time for the QMC with a \emph{parametric} initial quantum state;
and ii), as an immediate corollary,
be polynomial time for the QMC with a \emph{concrete} initial quantum state.
As a running example, the quantum IPv4 protocol is checked
to demonstrate the effectiveness of the proposed method.

Finally, we summarize the contributions of the paper as follows:
\begin{enumerate}
	\item a useful fidelity-based QCTL is presented;
	\item all BSCC subspaces are removed by their direct-sum,
	not individual ones,
	which makes our process more efficient than the existing work~\cite{FHT+17};
	\item the complexity is compatible/competitive
	when the QMC is provided with an initial quantum state, e.g. in~\cite{YYF13}.
\end{enumerate}

\paragraph{Organization of the paper}
Section~\ref{S2} gives the basic notions and notations from quantum computation.
Sections~\ref{S3} and \ref{S4} introduce the model of QMC and the logic of QCTL, respectively.
Section~\ref{S5} presents the model checking algorithm,
incorporating with an algebraic approach to the fidelity computation.
Section~\ref{S7} is the conclusion.
%Due to page limit,
%the detailed proofs are provided in Appendix~\ref{SS4},
%and an implementation of our method is discussed in Appendix~\ref{S6}.

\section{Preliminaries}\label{S2}
Here we recall some basic notions and notations in quantum computation.
Interested readers can refer to~\cite{NiC00,FYY13} for more details.

In this paper, we adopt the Dirac notations:
\begin{itemize}
	\item $\ket{\psi}$ stands for a unit column vector labelled with $\psi$;
	\item $\bra{\psi}:=\ket{\psi}^\dag$ is the Hermitian adjoint
	(i.e. complex conjugate and transpose) of $\ket{\psi}$;
	\item $\ip{\psi_1}{\psi_2}:=\bra{\psi_1} \ket{\psi_2}$
	is the inner product of $\ket{\psi_1}$ and $\ket{\psi_2}$; and
	\item $\op{\psi_1}{\psi_2}:=\ket{\psi_1} \otimes \bra{\psi_2}$ is the outer product,
	where $\otimes$ denotes tensor product.
\end{itemize}
Specifically, $\ket{i}$ with $i\in\mathbb{Z}^+$ denotes the vector,
in which the $i$th entry is $1$ and others are $0$.
Thus, $\ip{i}{j}=0$ holds for any positive integer $j \ne i$ by orthonormality.

Let $[n]$ ($n\in\mathbb{Z}^+$) denote the finite set $\{1,2,\ldots,n\}$.
Let $\h$ be a Hilbert space with dimension $d:=\dim(\h)$ throughout this paper.
Unit elements $\ket{\psi}$ of $\h$ are usually interpreted
as \emph{states} of a quantum system.
Since $\{\ket{i}: i\in[d]\}$ forms an orthonormal basis of $\h$,
any element $\ket{\psi}$ of $\h$
can be expressed as $\ket{\psi}=\sum_{i\in[d]} c_i\ket{i}$,
where $c_i \in \mathbb{C}$ ($i\in[d]$) satisfy $\sum_{i\in[d]} |c_i|^2=1$,
i.e. the quantum state $\ket{\psi}$ is entirely determined by those coefficients $c_i$.
In a product Hilbert space $\h \otimes \h'$,
let $\ket{\psi,\psi'}$ be a shorthand of
the product state $\ket{\psi}\ket{\psi'}:=\ket{\psi}\otimes\ket{\psi'}$
with $\ket{\psi} \in \h$ and $\ket{\psi'} \in \h'$.
%It is easy to validate that
%$\ip{\psi_1,\psi_1'}{\psi_2,\psi_2'}= \ip{\psi_1}{\psi_2}\ip{\psi_1'}{\psi_2'}$
%holds for any $\ket{\psi_1},\ket{\psi_2}$ in $\h$
%and $\ket{\psi_1'},\ket{\psi_2'}$ in $\h'$.
For any $\ket{\psi_1},\ket{\psi_2}$ in $\h$
and $\ket{\psi_1'},\ket{\psi_2'}$ in $\h'$,
the inner product of two product states $\ket{\psi_1,\psi'_1}$ and $\ket{\psi_2,\psi'_2}$
is defined by
$\ip{\psi_1,\psi_1'}{\psi_2,\psi_2'}=\ip{\psi_1}{\psi_2}\ip{\psi_1'}{\psi_2'}$.

Let $\LLL_\h$ be the set of linear operators on $\h$,
ranged over by letters in bold font, e.g. $\EE,\FF,\ind,\PP$.
For conciseness,
we will omit such a subscript $\h$ afterwards
if it is clear from the context.
A linear operator $\gamma$ is \emph{Hermitian} if $\gamma=\gamma^\dag$;
and it is \emph{positive}
if $\bra{\psi}\gamma\ket{\psi} \ge 0$ holds for any $\ket{\psi}\in\h$.
Given a Hermitian operator $\gamma$,
we have the spectral decomposition~\cite[Box~2.2]{NiC00} that
\begin{equation}
\gamma=\sum_{i\in[d]} \lambda_i \op{\psi_i}{\psi_i},
\end{equation}
where $\lambda_i \in \mathbb{R}$ ($i\in[d]$) are all eigenvalues of $\gamma$
and $\ket{\psi_i}$ are the corresponding eigenvectors.
The \emph{support} of $\gamma$ is the subspace of $\h$ spanned by
all eigenvectors associated with nonzero eigenvalues,
i.e. $\supp(\gamma):=\spn(\{\ket{\psi_i}:i\in[d] \wedge \lambda_i\ne 0\})
=\{\sum_{i\in[d]}c_i\ket{\psi_i}:c_i\in\mathbb{C} \wedge \lambda_i\ne 0\}$.
A \emph{projector} $\PP$ is a positive operator of
the form $\sum_{i\in[m]} \op{\psi_i}{\psi_i}$ with $m\le d$,
where $\ket{\psi_i}$ ($i\in[m]$) are orthonormal.
Obviously, there is a bijective map
between projectors $\PP=\sum_{i\in[m]} \op{\psi_i}{\psi_i}$
and subspaces of $\h$ that are spanned by $\{\ket{\psi_i}:i\in[m]\}$.
In sum, positive operators are Hermitian ones
whose eigenvalues are nonnegative;
and projectors are positive operators whose eigenvalues are $0$ or $1$.

The \emph{trace} of a linear operator $\gamma$ is defined as
$\tr(\gamma):=\sum_{i\in[d]} \bra{\psi_i}\gamma\ket{\psi_i}$
for any orthonormal basis $\{\ket{\psi_i}:i\in[d]\}$ of $\h$.
A \emph{density operator} (resp. \emph{partial density operator}) $\rho$ on $\h$
is a positive operator with trace $1$ (resp.~$\le 1$).
It gives rise to a generic way to describe quantum states:
if a density operator $\rho$ is $\op{\psi}{\psi}$ for some $\ket{\psi}\in \h$,
$\rho$ is said to be a \emph{pure} state;
otherwise it is a \emph{mixed} one,
i.e. $\rho=\sum_{i\in[d]} p_i \op{\psi_i}{\psi_i}$ under the spectral decomposition,
where $p_i$ ($i\in[d]$) are postive eigenvalues
(interpreted as the \emph{probabilities} of taking the pure states $\ket{\psi_i}$)
and together are $1$.
Sometimes,
the quantum states are described by the probabilistic ensemble form
$\{(p_i,\ket{\psi_i}):i\in[d]\}$
with $\ket{\psi_i}\in\h$ and $p_i\in\mathbb{R}^+$ satisfying $\sum_{i\in[d]} p_i=1$.
Note that the probabilistic ensemble form does not require
that $\ket{\psi_i}$ ($i\in[d]$) are orthogonal,
so it is more general.
Let $\DDD$ be the set of partial density operators on $\h$,
and $\DDD^1$ the set of density operators.
In a product Hilbert space $\h \otimes \h'$,
$\gamma \otimes \gamma'$ with $\gamma\in\LLL_\h$ and $\gamma'\in\LLL_{\h'}$
has the partial traces $\tr_{\h'}(\gamma \otimes \gamma'):=\tr(\gamma')\gamma$
and $\tr_\h(\gamma \otimes \gamma'):=\tr(\gamma)\gamma'$,
which result in linear operators in $\h$ and $\h'$, respectively.
The (partial) trace is defined to be linear in its input.

A \emph{super-operator} $\EEE$ on $\h$ is a linear operator on $\LLL_\h$,
ranged over by letters in calligraphic font, e.g. $\EEE,\FFF,\III,\PPP$.
A super-operator is \emph{completely positive}
if for any Hilbert space $\h'$,
the trivially extended operator $\EEE \otimes \III_{\h'}$
maps the set of positive operators on $\LLL_{\h \otimes \h'}$ to itself,
where $\III_{\h'}$ is the identity super-operator on $\h'$.
Let $\SSS$ be the set of completely positive super-operators on $\h$.
By Kraus representation~\cite[Thm.~8.3]{NiC00},
a super-operator $\EEE$ is completely positive on $\h$ if and only if
there are $m$ linear operators
$\EE_1,\EE_2,\ldots,\EE_m \in \LLL$ with $m \le d^2$ (called \emph{Kraus} operators),
such that for any $\gamma \in \LLL$, we have
\begin{equation}\label{eq:Kraus}
\EEE(\gamma) = \sum_{\ell\in[m]} \EE_\ell \,\gamma\, \EE_\ell^\dag.
\end{equation}
The description of $\EEE$ is given by those Kraus operators $\{\EE_\ell:\ell\in[m]\}$.
Thus, the sum $\EEE_1+\EEE_2$ of super-operators
$\EEE_1=\{\EE_{1,\ell}: \ell\in[m_1]\}$
and $\EEE_2=\{\EE_{2,\ell}: \ell\in[m_2]\}$
is given by the union
$\{\EE_{1,\ell}: \ell\in[m_1]\}\cup\{\EE_{2,\ell}: \ell\in[m_2]\}$;
and the composition $\EEE_2 \circ \EEE_1$ is given by
$\{\EE_{2,\ell_2}\EE_{1,\ell_1}: \ell_1\in[m_1] \wedge \ell_2\in[m_2]\}$.
In a product Hilbert space $\h \otimes \h'$,
for super-operators $\EEE=\{\EE_\ell:\ell\in[m]\} \in \SSS_\h$
and $\EEE'=\{\EE_\ell':\ell\in[m']\} \in \SSS_{\h'}$,
the product super-operator $\EEE\otimes\EEE'$ is given by
$\{\EE_\ell:\ell\in[m]\} \otimes \{\EE_\ell':\ell\in[m']\}
=\{\EE_\ell\otimes\EE_{\ell'}':\ell\in[m] \wedge \ell'\in[m']\}$.
It is easy to validate that
$\EEE\otimes\EEE'(\gamma\otimes\gamma')=\EEE(\gamma)\otimes\EEE'(\gamma')$
holds for any $\gamma\in\LLL_\h$ and $\gamma'\in\LLL_{\h'}$.

For a super-operator $\EEE\in\SSS$ and a density operator $\rho\in\DDD^1$,
the \emph{fidelity}\label{fid} is defined as
\begin{subequations}
\begin{equation}
	\fid(\EEE,\rho):=\tr\sqrt{\rho^{1/2} \EEE(\rho) \rho^{1/2}};
\end{equation}
and when $\rho$ is a pure state $\op{\psi}{\psi}$, it is simply
\begin{equation}
	\fid(\EEE,\op{\psi}{\psi}):=\sqrt{\bra{\psi} \EEE(\op{\psi}{\psi}) \ket{\psi}}.
\end{equation}
The fidelity reflects how well the quantum operation $\EEE$ has preserved
the quantum state $\rho$.
The better quantum state is preserved, the larger fidelity would be.
We can see $0 \le \fid(\EEE,\rho) \le 1$
where the equality in the first inquality holds
if and only if the supports of $\rho$ and $\EEE(\rho)$ are orthogonal,
and the equality in the second inequality holds if and only if $\EEE=\III$.
More technically, the fidelity measures the average angle
between the vectors in $\supp(\rho)$ and those in $\supp(\EEE(\rho))$,
which reveals that $\arccos\fid(\EEE,\rho)$ would be a standard metric
between $\rho$ and $\EEE(\rho)$. 
For conservation,
we would like to study the (minimum) fidelity of $\EEE$,
which is defined as
\begin{equation}
	\mfid(\EEE):=\min_{\rho \in \DDD^1} \fid(\EEE,\rho)
	=\min_{\ket{\psi} \in \h} \fid(\EEE,\op{\psi}{\psi}),
\end{equation}
\end{subequations}
where the last equation comes from the joint concavity~\cite[Ex.~9.19]{NiC00}.
%Finally we notice that
%$\mfid(\EEE_2 \circ \EEE_1) \ge \mfid(\EEE_1) \mfid(\EEE_2)$,
%which implies the fidelity does not admit the compositionality.

A trace pre-order $\lesssim$ can be defined on $\SSS$ as:
$\EEE_1 \lesssim \EEE_2$
if and only if $\tr(\EEE_1(\rho)) \le \tr(\EEE_2(\rho))$
holds for any $\rho \in \DDD$.
The equivalence $\EEE_1 \eqsim \EEE_2$ means
$\EEE_1 \lesssim \EEE_2$ and $\EEE_1 \gtrsim \EEE_2$.
For a super-operator $\EEE=\{\EE_\ell: \ell\in[m]\}$,
the completeness $\EEE \eqsim \III$ holds
if and only if $\sum_{\ell\in[m]} \EE_\ell^\dag \EE_\ell=\ind$
where $\ind$ is the identity operator.
Let $\SSS^{\lesssim \III}$ be the set of
\emph{trace-nonincreasing} super-operators $\EEE$,
i.e. $\SSS^{\lesssim \III}=\{\EEE \in \SSS:\EEE \lesssim \III\}$.
We would characterize quantum state evolution
by these super-operators $\EEE \in \SSS^{\lesssim \III}$ in the coming section.

\section{Quantum Markov Chain}\label{S3}
Let $AP$ be a set of atomic propositions throughout this paper.
\begin{definition}[{\cite[Def.~3.1]{FYY13}}]
	A labelled \emph{quantum Markov chain} (QMC for short) $\QMC$ over $\h$ is
	a triple $(S,Q,L)$, in which
	\begin{itemize}
		\item $S$ is a finite set of states,
		\item $Q: S\times S \to \SSS^{\lesssim \III}$
		is a transition super-operator matrix,
		satisfying $\sum_{t\in S} \linebreak[0] Q(s,t)\eqsim\III$ for each $s\in S$,
		and
		\item $L: S \to 2^{AP}$ is a labelling function.
	\end{itemize}
\end{definition}

Let $\ket{s}$ ($s \in S$) be the quantisation of classical state $s$,
and $\{\ket{s}: s\in S\}$ a set of orthonormal states serving as
the quantisation of classical system $S$.
Once all classical states in $S$ are ordered,
$\ket{s}$ denotes the $|S|$-dimensional vector,
in which the entry corresponding to $s$ is $1$ and others are $0$.
Further, $\h_\cq:=\mathcal{C} \otimes \h$
where $\mathcal{C}=\spn(\{\ket{s}: s\in S\})$
is the enlarged Hilbert space
corresponding to the whole classical--quantum system.
The dimension of $\h_\cq$ is $N:=nd$ where $n=|S|$.
In the QMC $\QMC$,
a state $\rho$ is given by a density operator on $\h_\cq$
with the mixed structure $\sum_{s\in S} \op{s}{s} \otimes \rho_s$
where $\rho_s \in \DDD$ ($s\in S$) satisfy $\sum_{s\in S} \tr(\rho_s)=1$.
The initial state is left \emph{parametric} in the model.

The transition super-operator matrix $Q$
is functionally analogous to the transition probability matrix
in an ordinary Markov chain (MC).
Actually, the former is more expressive than the latter,
and a QMC degenerates into an MC when $\h$ is one-dimensional.
Sometimes,
it is convenient to combine the super-operators in $Q$ together
to form a large single super-operator,
denoted $\FFF:=\sum_{s,t \in S} \{\op{t}{s}\} \otimes Q(s,t)$,
on $\h_\cq$.

A path $\omega$ in the QMC $\QMC$ is
an infinite state sequence in the form $s_0\,s_1\,s_2\cdots$,
where $s_i \in S$ and $Q(s_i,s_{i+1}) \ne 0$ for $i \ge 0$.
Let $\omega(i)$ be the $(i+1)$-th state of $\omega$ for $i \ge 0$,
e.g. $\omega(0)=s_0$ and $\omega(1)=s_1$ for $\omega=s_0\,s_1\,s_2\cdots$.
We denote by $Path(s)$ the set of all paths starting in $s$,
and by $Path_\textup{fin}(s)$
the set of all finite paths starting in $s$,
i.e. $Path_\textup{fin}(s):=\{\hat{\omega}:
\hat{\omega}\textup{ is a finite prefix of some }\omega\in Path(s)\}$.

\begin{example}[IPv4 protocol]\label{ex1}
	The IPv4 protocol~\cite{AHK04}
	aims at configuring IP addresses in a LAN of hosts.
	A quantum analogy goes as follows.
	When a new host joins in a LAN,
	it gets an IP address at random,
	encapsulated with its MAC address in the data message.
	Data messages are sent in quantum information, i.e. using density operators.
	The protocol determines whether the newly selected IP address is already in use
	by boardcasting a probe loading the message.
	If a host responds to the probe,
	which means that the IP address is occupied,
	the protocol updates the message with a new IP address.
	If no host responds within a unit of time,
	which may be caused by the noisy channel,
	the protocol repeats the probe by re-broadcasting the message.
	The current message is possibly corrupted by the noisy channel,
	but it is the only message we have,
	due to the no-cloning feature of quantum information~\cite{WoZ82}.
	If we do not get any response within a given time bound,
	the host would use the IP address chosen by the protocol.
	An error may be caused by missing probes.
	Finally,
	a server on the LAN records the new host's IP and MAC addresses in the message 
	after transferring through the noisy channel.
	Hence the fidelity between the initial MAC address and the final one
	is worth evaluating for the reliability of the channel.
	
	The QMC $\QMC_1=(S,Q,L)$ in Figure~\ref{fig:IPv4}
	describes the quantum IPv4 protocol.
	The state set $S$ is $\{s_0,s_1,s_2,s_3,s_4,s_5\}$,
	where $L(s_5)=\{\ok\}$, $L(s_4)=\{\error\}$,
	and other states are labelled with $\emptyset$.
	The state $s_0$ indicates that a new host joins in a LAN.
	The states $s_i$ ($i=1,2,3,4$) indicate although the address is occupied,
	no host responds the probe within $i$ units of time.
	If the total time $i=4$ does not run out, re-boardcasting a probe would take place,
	which leads to returning to the state $s_0$;
	otherwise the state $s_4$ indicates a wrong IP address configuration.
	The state $s_5$ indicates a proper IP address configuration.	
	The transition super-operator matrix $Q$ is given
	by the following nonzero entries in Kraus representation\footnote{%
		These super-operator entries are modelling noisy channels.
		In practice, each of them has a large proportion of being the indentity operator $\ind$
		with a small proportion of being noise operators,
		e.g. the bit flip $\mathbf{X}$ and the phase flip $\mathbf{Z}$,
		which would change density operators.
		However, to present our method concisely,
		we focus more on the situation where severe noises appear.
		Thus we set super-operator entries simply by those noise operators.}:
	\[
	\begin{aligned}
	& Q(s_0,s_1)=\{\op{1,+}{1,1}, \tfrac{4}{5}\op{1,-}{1,2}\}, 
	&& Q(s_0,s_5)=\{\tfrac{3}{5}\op{1,2}{1,2}, \op{2}{2}\otimes\ind\}, \\
	& Q(s_1,s_0)=\{\op{1,1}{1,+}, \tfrac{4}{5}\op{1,2}{1,-}\}, 
	&& Q(s_1,s_2)=\{\tfrac{3}{5}\op{1,2}{1,-}, \op{2}{2}\otimes\ind\}, \\
	&  Q(s_2,s_0)=\{\tfrac{12}{25}\,\mathbf{X}\otimes\ind,
	\tfrac{9}{25}\,\mathbf{X}\otimes\mathbf{X}\}, 
	&& Q(s_2,s_3)=\{\tfrac{16}{25}\,\ind\otimes\ind,
	\tfrac{12}{25}\,\ind\otimes\mathbf{X}\}, \\
	& Q(s_3,s_0)=\{\tfrac{12}{25}\,\ind\otimes\mathbf{Z},
	\tfrac{12}{25}\,\mathbf{Z}\otimes\ind\}, 
	&& Q(s_3,s_4)=\{\tfrac{16}{25}\,\ind\otimes\ind,
	\tfrac{9}{25}\,\mathbf{Z}\otimes\mathbf{Z}\}, \\
	& Q(s_4,s_4)=\{\ind\otimes\ind\}, 
	&& Q(s_5,s_5)=\{\ind\otimes\ind\},		
	\end{aligned}
	\] 
	where $\ket{\pm}=(\ket{1}\pm\ket{2})/\sqrt{2}$,
	$\ind=\op{1}{1}+\op{2}{2}$ is the identity operator,
	$\mathbf{X}=\op{1}{2}+\op{2}{1}$ is the bit flip
	and $\mathbf{Z}=\op{1}{1}-\op{2}{2}$ is the phase flip.
	It is easy to validate that 
	$\sum_{t\in S} Q(s,t)\eqsim\III$ holds for each $s\in S$.
	
	We can combine all these super-operators on $\h$
	as a single super-operator on $\h_\cq$:
	\[
	\begin{aligned}
	\FFF = \ & \{\op{s_1}{s_0}\} \otimes Q(s_0,s_1)
	+\{\op{s_5}{s_0}\} \otimes Q(s_0,s_5) + 
	\{\op{s_0}{s_1}\} \otimes Q(s_1,s_0) \ + \\
	& \{\op{s_2}{s_1}\} \otimes Q(s_1,s_2)
	+\{\op{s_0}{s_2}\} \otimes Q(s_2,s_0)
	+\{\op{s_3}{s_2}\} \otimes Q(s_2,s_3) \ + \\
	& \{\op{s_0}{s_3}\} \otimes Q(s_3,s_0)
	+\{\op{s_4}{s_3}\} \otimes Q(s_3,s_4)
	+\{\op{s_4}{s_4}\} \otimes \III
	+\{\op{s_5}{s_5}\} \otimes \III,
	\end{aligned}
	\]
	in which the left operand of the tensor product in a term
	is a super-operator on $\mathcal{C}$
	and the right operand is a super-operator on $\h$.
	
		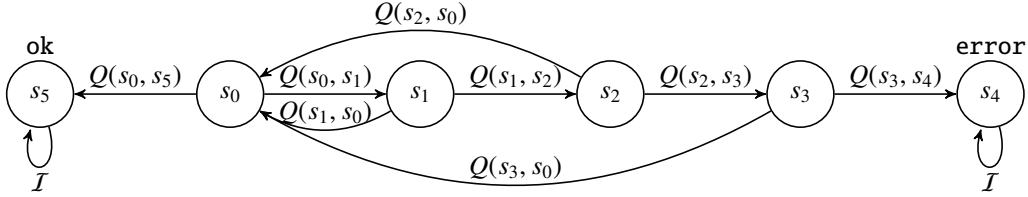
\begin{figure}[H]
		\centering
		\begin{tikzpicture}[->,>=stealth',auto,node distance=2.5cm,semithick,inner sep=2pt]
		\node[state](s0){$s_0$};
		\node[state](s1)[right of=s0]{$s_1$};
		\node[state](s2)[right of=s1]{$s_2$};
		\node[state](s3)[right of=s2]{$s_3$};
		\node[state,label=above:$\error$](s4)[right of=s3]{$s_4$};
		\node[state,label=above:$\ok$](s5)[left of=s0]{$s_5$};
		
		\draw(s0)edge[]node{$Q(s_0,s_1)$}(s1);
		\draw(s1)edge[bend left]node[above]{$Q(s_1,s_0)$}(s0);
		\draw(s1)edge[]node{$Q(s_1,s_2)$}(s2);
		\draw(s2)edge[bend right]node[above]{$Q(s_2,s_0)$}(s0);
		\draw(s2)edge[]node{$Q(s_2,s_3)$}(s3);
		\draw(s3)edge[bend left]node[above]{$Q(s_3,s_0)$}(s0);
		\draw(s3)edge[]node{$Q(s_3,s_4)$}(s4);
		\draw(s0)edge[]node[above]{$Q(s_0,s_5)$}(s5);
		\draw(s4)edge[loop below]node{$\III$}(s4);
		\draw(s5)edge[loop below]node{$\III$}(s5);
		\end{tikzpicture}
		\caption{QMC for IPv4 protocol}\label{fig:IPv4}
	\end{figure}
	
	In the QMC $\QMC_1$,
	$\omega_1=s_3\,s_0\,s_1\,s_0\,s_5\,s_5\cdots$ is a path starting in $s_3$,
	where $\omega_1(0)=s_3$,
	$\omega_1(1)=\omega_1(3)=s_0$,
	$\omega_1(2)=s_1$,
	and $\omega_1(i)=s_5$ for $i \ge 4$;
	while $\hat{\omega}_1=s_3\,s_0\,s_1\,s_0\,s_5$ is a finite prefix of $\omega_1$.
	Therefore, we have $\omega_1 \in Path(s_3)$
	and $\hat{\omega}_1 \in Path_{\textup{fin}}(s_3)$.
\end{example}

To reason about quantitative properties of QMC,
a super-operator valued measure (SOVM) space over paths
could be established as follows.
Recall that:
\begin{definition}
	A \emph{measurable space} is a pair $(\Omega,\Sigma)$,
	where $\Omega$ is a nonempty set
	and $\Sigma$ is a $\sigma$-algebra on $\Omega$;
	in addition an \emph{SOVM space} is a triple $(\Omega,\Sigma,\Delta)$,
	where $(\Omega,\Sigma)$ is a measurable space
	and $\Delta: \Sigma \to \SSS^{\lesssim \III}$ is an SOVM,
	satisfying:
	\begin{itemize}
		\item $\Delta(\Omega) \eqsim \III$, and
		\item $\Delta(\biguplus_i A_i) \eqsim \sum_i \Delta(A_i)$
		for any pairwise disjoint $A_i \in \Sigma$.
	\end{itemize}
\end{definition}

For a given finite path $\hat{\omega} \in Path_\textup{fin}(s)$,
we define the cylinder set as
\begin{equation}\label{eq:cylinder}
Cyl(\hat{\omega}):=
\{\omega \in Path(s): \omega\textup{ has the prefix }\hat{\omega}\};
\end{equation}
and for $B \subseteq Path_\textup{fin}(s)$,
we extend~\eqref{eq:cylinder} by
$Cyl(B):=\bigcup_{\hat{\omega}\in B} Cyl(\hat{\omega})$.
Particularly, we have $Cyl(s)=Path(s)$.
Let $\Omega=Path(s)$ for an appointed $s\in S$,
and $\Pi \subseteq 2^\Omega$ be the countable set of all cylinder sets
$\{Cyl(\hat{\omega}): \hat{\omega}\in Path_\textup{fin}(s)\}$
plus the emptyset $\emptyset$.
By~\cite[Chapt.~10]{BaK08},
there is a smallest $\sigma$-algebra $\Sigma$ of $\Pi$
that contains $\Pi$
and is closed under countable union and complement.
It is clear that the pair $(\Omega,\Sigma)$ forms a measurable space.

Next, for a given finite path $\hat{\omega}=s_0\,s_1 \cdots s_n$,
we define the accumulated super-operator along with $\hat{\omega}$ as
\begin{equation}
\Delta(Cyl(\hat{\omega})):=\left\{
\begin{array}{ll}
\III & \quad \textup{if } n=0, \\
Q(s_{n-1},s_n)\circ\cdots\circ Q(s_0,s_1) & \quad \textup{otherwise}.
\end{array} \right.
\end{equation}
By~\cite[Thm.~3.2]{FYY13},
the domain of $\Delta$ can be extended to $\Sigma$,
i.e. $\Delta:\Sigma\to\SSS^{\lesssim \III}$,
which is unique under the countable union $\bigcup_i A_i$ for any $A_i \in \Pi$
and is an equivalence class of super-operators
in terms of $\eqsim$ under the complement $A^\textup{c}$ for some $A \in \Pi$.
Hence the triple $(\Omega,\Sigma,\Delta)$ forms an SOVM space.

\section{Quantum Computation Tree Logic}\label{S4}
We now introduce a quantum extension of computation tree logic (QCTL).
The basic idea is to
replace the probability measure in the logic of~\cite{FYY13} with fidelity.
As we mentioned in the introduction,
fidelity is useful in comparing quantum states.
In practice, the preparation of any quantum state is limited
by imperfections and noises,
and one often needs to find out how close the produced state is to the intended one.
In many occasions,
fidelity can detect the effect of a noisy channel but probability measure cannot.

In the following,
we present the syntax and semantics of the new logic,
then compare it with probabilistic CTL (PCTL)~\cite{HaJ89}
and with the QCTL presented in~\cite{FYY13}.

\begin{definition}
	The syntax of QCTL consists of
	the state formulas $\Phi$ and path formulas $\phi$:
	\[
	\begin{aligned}
	\Phi &:= \textup{a} \ |\ \neg\Phi \ |\ \Phi_1 \wedge \Phi_2
	\ |\ \QQ_{\sim \tau}[\phi] \\
	\phi &:= \nxt \Phi \ |\ \Phi_1 \ntl^{\le k} \Phi_2 \ |\ \Phi_1 \ntl \Phi_2
	\end{aligned}
	\]
	where $\textup{a}\in AP$ is an atomic proposition,
	$\sim\,\in \{<,\le,=,\ge,>,\ne\}$ is a comparison operator,
	$\tau \in \mathbb{Q} \cap [0,1]$ is a threshold,
	and $k \ge 0$ is a step bound.
\end{definition}
The state formula $\QQ_{\sim \tau}[\phi]$ in QCTL is called
the \emph{fidelity-quantifier} formula,
and other state formulas are \emph{basic} ones.
The three kinds of path formulas $\nxt \Phi$,
$\Phi_1 \ntl^{\le k} \Phi_2$ and $\Phi_1 \ntl \Phi_2$ are
the \emph{next},
the \emph{bounded-until}
and the \emph{unbounded-until} formulas, respectively.

\begin{definition}
	The semantics of QCTL interpreted over a QMC $\QMC=(S,Q,L)$
	is given by the satisfaction relation $\models$:
	\[
	\begin{aligned}
	s & \models \textup{a}
	&& \textup{if } \textup{a} \in L(s), \\
	s & \models \neg \Phi
	&& \textup{if } s \not\models \Phi, \\
	s & \models \Phi_1 \wedge \Phi_2
	&& \textup{if } s \models \Phi_1 \wedge s \models \Phi_2, \\
	s & \models \QQ_{\sim \tau}[\phi]
	&& \textup{if } \mfid(\Delta(\{\omega\in Path(s): \omega\models\phi\}))\sim\tau, \\
	\omega & \models \nxt \Phi
	&& \textup{if } \omega(1) \models \Phi, \\
	\omega & \models \Phi_1 \ntl^{\le k} \Phi_2
	&& \textup{if } \exists\, i\le k:
	(\omega(i) \models \Phi_2 \wedge \forall\, j<i: \omega(j) \models \Phi_1), \\
	\omega & \models \Phi_1 \ntl \Phi_2
	&& \textup{if } \exists\, i:
	(\omega(i) \models \Phi_2 \wedge \forall\, j<i: \omega(j) \models \Phi_1).
	\end{aligned}
	\]
	Other logic connectives $\vee$, $\rightarrow$ and $\leftrightarrow$
	can be easily derived by $\neg$ and $\wedge$ as usual. 
\end{definition}

For any path formula $\phi$,
the path set $A=\{\omega\in Path(s): \omega\models\phi\}$ is measurable,
since
\begin{itemize}
	\item if $\phi=\nxt \Phi$,
	$A$ is the finite union of those cylinder sets $Cyl(s\,t)$
	that satisfy $t\models\Phi$;
	\item if $\phi=\Phi_1 \ntl^{\le k} \Phi_2$,
	$A$ is the finite union of $Cyl(s_0 \cdots s_i)$
	for some $i\le k$,
	that satisfy $s_0=s$, $s_i\models \Phi_2$,
	and $s_j\models \Phi_1$ for each $j<i$;
	and
	\item if $\phi=\Phi_1 \ntl \Phi_2$,
	$A$ is the countable union of $Cyl(s_0 \cdots s_i)$
	for some $i\ge 0$,
	that satisfy $s_0=s$, $s_i\models \Phi_2$,
	and $s_j\models \Phi_1$ for each $j<i$.
\end{itemize}
Thereby, the set $A$ belongs to the $\sigma$-algebra $\Sigma$
and in particular is a countable union of cylinder sets,
which entails that the SOVM $\Delta(A)$ is uniquely defined.
For conciseness,
we will write $\Delta(\hat{\omega})$ for $\Delta(Cyl(\hat{\omega}))$
and $\Delta(\phi)$ for $\Delta(\{\omega\in Path(s): \omega\models\phi\})$
afterwards.

%Sometimes we will consider a variant of path formula $\Phi_1 \mathbf{U}^{[j,k]} \Phi_2$.
%A path $\omega \models \Phi_1 \mathbf{U}^{[j,k]} \Phi_2$ iff $\exists i \in [j,k]$ s.t.
%$\forall l \in [j,i), \omega(l) \models \Phi_1$
%and $\forall l \in [i,k], \omega(l) \models \Phi_2$.
\begin{example}\label{ex2}
	From the QMC $\QMC_1$ together with
	the path $\omega_1=s_3\,s_0\,s_1\,s_0\,s_5\,s_5\cdots$
	shown in Example~\ref{ex1},
	we can see
	\begin{itemize}
		\item $s_5 \models \ok$
		and $s \not\models \ok$ for each $s\in S\setminus\{s_5\}$;
		\item $\omega_1 \models \true\, \ntl \ok$,
		as $\omega_1(4) \models \ok$
		and $\omega_1(j) \models \true$ for each $j<4$.
	\end{itemize}
	For each $s\in S$, we can establish an SOVM space $(\Omega,\Sigma,\Delta)$
	over the path set $Path(s)$ of $\QMC_1$.
	To demonstrate the generality of the method developed in this paper,
	we choose $\Omega=Path(s_3)$.
	The SOVM $\Delta(\hat{\omega}_1)$ is calculated as
	\[
	\begin{aligned}
	\Delta(\hat{\omega}_1)=\Delta(Cyl(\hat{\omega}_1))
	& =Q(s_0,s_5)\circ Q(s_1,s_0)\circ Q(s_0,s_1)\circ Q(s_3,s_0) \\
	& =Q(s_0,s_5)\circ Q(s_1,s_0)\circ Q(s_0,s_1)\circ 
	\{\tfrac{12}{25}\,\ind\otimes\mathbf{Z},\tfrac{12}{25}\,\mathbf{Z}\otimes\ind\} \\
	& =Q(s_0,s_5)\circ Q(s_1,s_0)\circ
	\{\tfrac{12\sqrt{2}}{25}\op{1,+}{1,1},\tfrac{48\sqrt{2}}{125}\op{1,-}{1,2}\} \\
	& =Q(s_0,s_5)\circ\{\tfrac{12\sqrt{2}}{25}\op{1,1}{1,1},
	\tfrac{192\sqrt{2}}{625}\op{1,2}{1,2}\} \\
	& =\{\tfrac{576\sqrt{2}}{3125}\op{1,2}{1,2}\}.
	\end{aligned}
	\]
	In details,
	we calculate the composition of super-operators using right associativity law,
	e.g.
	\[
	\begin{aligned}
	& Q(s_0,s_1)\circ Q(s_3,s_0) \\
	=\ & \{\op{1,+}{1,1}, \tfrac{4}{5}\op{1,-}{1,2}\}\circ 
	\{\tfrac{12}{25}\,\ind\otimes\mathbf{Z},\tfrac{12}{25}\,\mathbf{Z}\otimes\ind\} \\
	=\ & \{\tfrac{12}{25}\op{1,+}{1,1} (\ind\otimes\mathbf{Z}),
	\tfrac{12}{25}\op{1,+}{1,1} (\mathbf{Z}\otimes\ind),
	\tfrac{48}{125}\op{1,-}{1,2} (\ind\otimes\mathbf{Z}),
	\tfrac{48}{125}\op{1,-}{1,2} (\mathbf{Z}\otimes\ind)\} \\
	=\ & \{\tfrac{12}{25}\ket{1,+} (\bra{1}\otimes\bra{1}) (\ind\otimes\mathbf{Z}),
	\tfrac{12}{25}\ket{1,+} (\bra{1}\otimes\bra{1}) (\mathbf{Z}\otimes\ind),
	\tfrac{48}{125}\ket{1,-} (\bra{1}\otimes\bra{2}) (\ind\otimes\mathbf{Z}), \\
	& \quad \tfrac{48}{125}\ket{1,-} (\bra{1}\otimes\bra{2}) (\mathbf{Z}\otimes\ind)\} \\
	=\ & \{\tfrac{12}{25}\ket{1,+} [(\bra{1}\ind)\otimes(\bra{1}\mathbf{Z})],
	\tfrac{12}{25}\ket{1,+} [(\bra{1}\mathbf{Z})\otimes(\bra{1}\ind)],
	\tfrac{48}{125}\ket{1,-} [(\bra{1}\ind)\otimes(\bra{2}\mathbf{Z})], \\
	& \quad \tfrac{48}{125}\ket{1,-} [(\bra{1}\mathbf{Z})\otimes(\bra{2}\ind)]\} \\
	=\ & \{\tfrac{12}{25}\ket{1,+} (\bra{1}\otimes\bra{1}),
	\tfrac{12}{25}\ket{1,+} (\bra{1}\otimes\bra{1}),
	\tfrac{48}{125}\ket{1,-} (-\bra{1}\otimes\bra{2}),
	\tfrac{48}{125}\ket{1,-} (\bra{1}\otimes\bra{2})\} \\
	=\ & \{\tfrac{12}{25}\op{1,+}{1,1},
	\tfrac{12}{25}\op{1,+}{1,1},
	-\tfrac{48}{125}\op{1,-}{1,2},
	\tfrac{48}{125}\op{1,-}{1,2}\} \\
	=\ & \{\tfrac{12\sqrt{2}}{25}\op{1,+}{1,1},\tfrac{48\sqrt{2}}{125}\op{1,-}{1,2}\},
	\end{aligned}
	\]
	where the last equation follows from the fact
	that they are two Kraus representations of a super-operator.
	Since $\omega_1 \in \Omega$
	and $\omega_1 \models \true\, \ntl \ok$,
	we have the lower bound $\Delta(\true\, \ntl \ok)
	\gtrsim \{\tfrac{576\sqrt{2}}{3125}\op{1,2}{1,2}\}$.
\end{example}

Finally, we point out the difference
between the PCTL in~\cite{HaJ89}, the QCTL in~\cite{FYY13} and our QCTL.
The PCTL extends CTL by
introducing a probability-quantifier $\mathfrak{P}_{\le\tau}(\phi)$
that compares the probability of the measurable event specified by $\phi$
with the threshold $\tau$;
and decides it over a MC with a specific initial state (probability distribution).
The QCTL in~\cite{FYY13} introduces
an SOVM-quantifier $\mathfrak{Q}_{\lesssim \EEE}(\phi)$
that compares the SOVM on $\phi$ with the super-operator threshold $\EEE$
under the trace pre-order $\lesssim$;
and decides it over a QMC with a specific initial state (density operator).
Whereas, ours introduces a fidelity-quantifier $\QQ_{\le\tau}(\phi)$
that compares the fidelity of the SOVM on $\phi$ with the threshold $\tau$;
and aims to decide it over a QMC with a parametric initial state.
The parametric model is more expressive,
and thus our method would be potentially more appliable.
How to consider the SOVM-quantifier on a parametric QMC would be one of our future work.

\section{Model Checking Algorithm}\label{S5}
In this section, we present the model checking algorithm
for a given QMC $\QMC=(S,Q,L)$ and a QCTL state formula $\Phi$.
The algorithm would decide $s\models\Phi$ for an appointed state $s\in S$,
or equivalently compute the set of all states satisfying $\Phi$,
i.e. $Sat(\Phi):=\{s\in S: s\models \Phi\}$.
Since the definition of QCTL is mutually inductive,
this goal will be reached in three steps:
\begin{enumerate}
	\item deciding basic state formulas
	(except for the fidelity-quantifier one),
	\item synthesizing the super-operators of path formulas, and
	\item deciding the fidelity-quantifier formula.
\end{enumerate}
%The first step can be achieved in polynomial time by standard literature~\cite{HaJ89},
%thus is omitted.

\subsection{Deciding basic state formulas}
For basic state formulas,
the satisfying sets are calculated by their definitions:
\begin{itemize}
	\item $Sat(\textup{a})=\{s\in S: \textup{a}\in L(s)\}$;
	\item $Sat(\neg\Phi)=S \setminus Sat(\Phi)$,
	provided that $Sat(\Phi)$ is known; and
	\item $Sat(\Phi_1\wedge\Phi_2)=Sat(\Phi_1) \cap Sat(\Phi_2)$,
	provided that $Sat(\Phi_1)$ and $Sat(\Phi_2)$ are known.
\end{itemize}
Obviously,
the top-level logic connective of those formulas
requires merely a scan over the labelling function $L$ on $S$,
which is in $\BigO(n)$.
Hence, deciding basic state formulas is linear time
w.r.t. the size of $\QMC$.

\begin{example}
	From the QMC $\QMC_1$ shown in Example~\ref{ex1},
	it is easy to calculate
	\begin{itemize}
		\item $Sat(\ok)=\{s_5\}$, $Sat(\error)=\{s_4\}$;
		\item $Sat(\neg\ok)=S\setminus Sat(\ok)
		=\{s_0,s_1,s_2,s_3,s_4\}$;
		\item $Sat(\neg\error)=S\setminus Sat(\error)
		=\{s_0,s_1,s_2,s_3,s_5\}$; and
		\item $Sat(\neg(\ok\vee\error))
		=Sat(\neg\ok\wedge\neg\error)
		=Sat(\neg\ok) \cap Sat(\neg\error)
		=\{s_0,s_1,s_2,s_3\}$.
	\end{itemize}
\end{example}

\subsection{Synthesizing the super-operators of path formulas}
Let $\PPP_s$ denote the projection super-operator
$\{\op{s}{s}\} \otimes \III=\{\op{s}{s}\otimes\ind\}$
on the enlarged Hilbert space $\h_\cq$,
and $\PPP_\Phi:=\{\sum_{s\models\Phi} \op{s}{s}\} \otimes \III
=\{\sum_{s\models\Phi}\op{s}{s}\otimes\ind\}$.
Utilizing the mixed structure of the classical--quantum state
$\rho=\sum_{s\in S} \op{s}{s} \otimes \rho_s$,
we have the nice property
\begin{equation}
	\rho=\sum_{s\models\Phi} \op{s}{s} \otimes \rho_s
	+\sum_{s\models\neg\Phi} \op{s}{s} \otimes \rho_s
	=\PPP_\Phi(\rho)+\PPP_{\neg\Phi}(\rho)
\end{equation}
So, fixing an initial classical state $s$,
we can obtain the SOVMs of path formulas as follows.

\begin{itemize}
	\begin{subequations}\label{eq:paths}
	\item Supposing that $Sat(\Phi)$ is known, we have
	\begin{equation}
		\Delta(\nxt \Phi)
		=\Delta\left(\biguplus_{t\models\Phi}Cyl(s\,t)\right)
		=\sum_{t\models\Phi} \Delta(s\,t)
		=\sum_{t\models\Phi} Q(s,t),
	\end{equation}
	where $\uplus$ denotes disjoint union.
	\item Supposing that $Sat(\Phi_1)$ and $Sat(\Phi_2)$ are known, we have
	\begin{align}
		\Delta(\Phi_1 \ntl^{\le k} \Phi_2)
		&= \Delta\left(\biguplus_{i=0}^k \left\{\omega\in Path(s):
		\omega(i)\models\Phi_2 \wedge \bigwedge_{j=0}^{i-1}
		\omega(j)\models\Phi_1\wedge\neg\Phi_2 \right\}\right) \notag \\
		&= \sum_{i=0}^k \Delta\left(\left\{\omega\in Path(s):
		\omega(i)\models\Phi_2 \wedge \bigwedge_{j=0}^{i-1}
		\omega(j)\models\Phi_1\wedge\neg\Phi_2 \right\}\right) \notag \\
		&= \sum_{i=0}^k \tr_\mathcal{C}(\PPP_{\Phi_2} \circ
		(\FFF \circ \PPP_{\Phi_1\wedge\neg\Phi_2})^i \circ \PPP_s),
	\end{align}
	where $\tr_\mathcal{C}=\{\bra{s}\otimes\ind:s\in S\}$ is the partial trace
	that traces out the classical system $\mathcal{C}$
	and $\FFF=\sum_{s,t \in S} \{\op{t}{s}\} \otimes Q(s,t)$
	is defined in Section~\ref{S3}.
	
	\item Supposing that $Sat(\Phi_1)$ and $Sat(\Phi_2)$ are known, we have
	\begin{align}
		\Delta(\Phi_1 \ntl \Phi_2)
		&= \Delta\left(\biguplus_{i=0}^\infty \left\{\omega\in Path(s):
		\omega(i)\models\Phi_2 \wedge \bigwedge_{j=0}^{i-1}
		\omega(j)\models\Phi_1\wedge\neg\Phi_2 \right\}\right) \notag \\
		&= \sum_{i=0}^\infty \Delta\left(\left\{\omega\in Path(s):
		\omega(i)\models\Phi_2 \wedge \bigwedge_{j=0}^{i-1}
		\omega(j)\models\Phi_1\wedge\neg\Phi_2 \right\}\right) \notag \\
		\label{eq:ntl} &= \sum_{i=0}^\infty \tr_\mathcal{C}(\PPP_{\Phi_2} \circ
		(\FFF \circ \PPP_{\Phi_1\wedge\neg\Phi_2})^i \circ \PPP_s).
	\end{align}
	\end{subequations}
\end{itemize}
For the latter two cases,
we classify all satisfying paths $\omega$ upon the first timestamp $i$
that satisfies $\omega(i)\models\Phi_2$ and $\omega(j)\models\Phi_1$ for each $j<i$
(or equivalently the unique timestamp $i$
that satisfies $\omega(i)\models\Phi_2$
and $\omega(j)\models\Phi_1\wedge\neg\Phi_2$ for each $j<i$).
Thereby, the resulting sets $A_i=\{\omega\in Path(s):
\omega(i)\models\Phi_2 \wedge \bigwedge_{j=0}^{i-1}
\omega(j)\models\Phi_1\wedge\neg\Phi_2\}$ are pairwise disjont,
and their SOVMs are obtained as $\tr_\mathcal{C}(\PPP_{\Phi_2} \circ
(\FFF \circ \PPP_{\Phi_1\wedge\neg\Phi_2})^i \circ \PPP_s)$.

We notice that all super-operators,
say $\FFF \circ \PPP_{\Phi_1\wedge\neg\Phi_2}$,
in the SOVMs~\eqref{eq:paths} has the property
$\FFF \circ \PPP_{\Phi_1\wedge\neg\Phi_2}
=\FFF \circ \PPP_{\Phi_1\wedge\neg\Phi_2} \circ \sum_{s\in S} \PPP_s$,
which implies all density operators $\rho\in\DDD_{\h_\cq}$
occurring in our analysis keep
the mixed structure $\sum_{s\in S} \op{s}{s}\otimes\rho_s$ with $\rho_s\in\DDD$.

\begin{example}\label{ex4}
	Under the SOVM space $(\Omega,\Sigma,\Delta)$ established in Example~\ref{ex2},
	we consider the path formula $\phi_1=\true\,\ntl \ok$.
	The satisfying path sets are disjoint
	$A_i=\{\omega\in \Omega:
	\omega(i)\models\ok \wedge \bigwedge_{j=0}^{i-1}
	\omega(j)\models\neg\ok\}$ ($i\ge 0$);
	and their SOVMs are:
	\[
	\begin{aligned}
		\Delta(A_0) &= \tr_\mathcal{C}(\PPP_\ok \circ \PPP_{s_3})	=0, \\
		\Delta(A_1) &= \tr_\mathcal{C}(\PPP_\ok \circ
		(\FFF \circ \PPP_{\neg\ok}) \circ \PPP_{s_3}) =0, \\
		\Delta(A_2) &= \tr_\mathcal{C}(\PPP_\ok \circ
		(\FFF \circ \PPP_{\neg\ok})^2 \circ \PPP_{s_3})
		=\{ \tfrac{36\sqrt{2}}{125}\op{1,2}{1,2},
		\tfrac{12}{25}\op{2}{2}\otimes\mathbf{Z},
		\tfrac{12}{25}\op{2}{2}\otimes\ind \}, \\
		\Delta(A_3) &= \tr_\mathcal{C}(\PPP_\ok \circ
		(\FFF \circ \PPP_{\neg\ok})^3 \circ \PPP_{s_3})	=0, \\
		\Delta(A_4) &= \tr_\mathcal{C}(\PPP_\ok \circ
		(\FFF \circ \PPP_{\neg\ok})^4 \circ \PPP_{s_3})
		=\{ \tfrac{576\sqrt{2}}{3125}\op{1,2}{1,2}\}, \\
		\Delta(A_5) &= \tr_\mathcal{C}(\PPP_\ok \circ
		(\FFF \circ \PPP_{\neg\ok})^5 \circ \PPP_{s_3}) 
		=\{ \tfrac{1728\sqrt{2}}{15625}\op{2,2}{1,2},
		\tfrac{1296\sqrt{2}}{15625}\op{2,1}{1,2} \},                                                                                                                                                                                                                                                                                                                                                                                                                                                                                                                                                                                                                                                                                                                                                            
	\end{aligned}
	\]
	and so on.
\end{example}

Although the three kinds of path formulas have the SOVMs~\eqref{eq:paths},
the super-operators are not expressed in an explicit form,
i.e. there are too many Kraus operators to
make up the super-operators $\Delta(\phi)$.
In particular, $\Delta(\Phi_1 \ntl \Phi_2)$ in~\eqref{eq:ntl}
is even not expressed in a closed form.
In the following,
we will construct explicit matrix representations for these super-operators,
particularly for $\Delta(\Phi_1 \ntl \Phi_2)$.
%Using the notations:
%\begin{itemize}
%	\item for a linear operator $\gamma$,
%	the function $\mv(\gamma):=(\gamma \otimes \ind) \sum_{l \in [d]} \ket{l,l}$
%	rearranges entries of $\gamma$ as a column vector, and
%	\item for a super-operator $\EEE=\{\EE_\ell: \ell\in[m]\}$,
%	the matrix representation~\cite[Def.~2.2]{YYF13} is defined as
%	$\sm(\EEE):=\sum_{\ell \in [m]} \EE_\ell \otimes \EE_\ell^*$
%	where $*$ denotes complex conjugate,
%\end{itemize}
%it is easy to validate $\mv(\EEE(\gamma))=\sm(\EEE)\mv(\gamma)$
%and $\sm(\EEE_2 \circ \EEE_1)=\sm(\EEE_2) \sm(\EEE_1)$.
A natural idea is:
\begin{enumerate}
	\item using the matrix representation of $\Delta(\Phi_1 \ntl \Phi_2)$,
	which is analogous to a geometric series
	with common ratio---the matrix representation of
	$\FFF_{\Phi_1\wedge\neg\Phi_2}:=\FFF \circ \PPP_{\Phi_1\wedge\neg\Phi_2}$;
	and
	\item reformulating it as a matrix fraction.
\end{enumerate}

However,
$\FFF_{\Phi_1\wedge\neg\Phi_2}$ may have some fixed-point $\gamma$
(or equivalently the matrix representation of $\FFF_{\Phi_1\wedge\neg\Phi_2}$
may have eigenvalue $1$),
which makes the matrix fraction divergent.
To overcome the trouble, inspired by~\cite{FHT+17},
we will remove the bottom strongly connected component (BSCC) subspaces $\Gamma$
that cover all fixed-points $\gamma$ of $\FFF_{\Phi_1\wedge\neg\Phi_2}$,
i.e. $\supp(\gamma) \subseteq \Gamma$.
Recall that:
\begin{definition}
	For a super-operator $\EEE\in\SSS$,
	a subspace $\Gamma$ of $\h$ is \emph{bottom}
	if for any pure state $\ket{\psi}\in\Gamma$,
	the support of $\EEE(\op{\psi}{\psi})$ is contained in $\Gamma$;
	it is \emph{SCC}
	if for any pure states $\ket{\psi_1},\ket{\psi_2}\in\Gamma$,
	$\ket{\psi_2}$ is in
	$\spn(\bigcup_{i=0}^\infty \supp(\EEE^i(\op{\psi_1}{\psi_1})))$;
	and it is \emph{BSCC} if it is bottom and SCC.
\end{definition}

We characterize the fixed-point of $\FFF_{\Phi_1\wedge\neg\Phi_2}$
by the stationary equation
\begin{equation}\label{eq:bottom}
	\FFF_{\Phi_1\wedge\neg\Phi_2}(\gamma)=\gamma
	\qquad (\gamma=\gamma^\dag\in\LLL_{\h_\cq}),
\end{equation}
where $\gamma$ are unknown variables
and $\FFF_{\Phi_1\wedge\neg\Phi_2}$ gives rise to coefficients.
It is a system of homogeneous linear equations.
Let $\gamma_i$ ($i\in[m]$)
be all linearly independent solutions of~\eqref{eq:bottom}.
Thanks to the property $\FFF_{\Phi_1\wedge\neg\Phi_2}
=\FFF_{\Phi_1\wedge\neg\Phi_2} \circ \sum_{s\in S}\PPP_s$,
the number of real variables in the Hermitian operator $\gamma$
can be bounded by $nd^2$.
So the number $m$ of these solutions is also bounded by $nd^2$.
We proceed to find out the BSCC subspaces by the following lemma.

\begin{lemma}\label{lem:all}
	The direct-sum of all BSCC subspaces w.r.t. $\FFF_{\Phi_1\wedge\neg\Phi_2}$
	is $\spn(\bigcup_{i\in[m]} \supp(\gamma_i))$.
\end{lemma}
\begin{proof}
	We first prove $\Gamma:=\spn(\bigcup_{i\in[m]} \supp(\gamma_i))$
	is the direct-sum of \emph{some} BSCC subspaces
	that covers all fixed-point of $\FFF_{\Phi_1\wedge\neg\Phi_2}$;
	then show it is the direct-sum of \emph{all} BSCC subspaces.
	
	Let $\gamma_i=\sum_{j\in[N]} \lambda_{i,j} \op{\Psi_{i,j}}{\Psi_{i,j}}$
	be the spectral decomposition of $\gamma_i$,
	where $\lambda_{i,j} \in \mathbb{R}$ ($j\in[N]$) are all eigenvalues of $\gamma_i$
	and $\ket{\Psi_{i,j}}$ are the corresponding eigenvectors.
	Define
	\[
	\begin{aligned}
	\gamma_i^+ &:=\sum \{|\, \lambda_{i,j} \op{\Psi_{i,j}}{\Psi_{i,j}}:
	j\in[N] \wedge \lambda_{i,j}>0 \,|\} \\
	\gamma_i^- &:=\sum \{|\, \lambda_{i,j} \op{\Psi_{i,j}}{\Psi_{i,j}}:
	j\in[N] \wedge \lambda_{i,j}<0 \,|\},
	\end{aligned}
	\]
	where $\{|\,\cdot\,|\}$ denotes a multiset,
	as the positive and the negative parts of $\gamma_i$, respectively.
	Utilizing the fact that $\FFF_{\Phi_1\wedge\neg\Phi_2}$ is completely positive,
	the positive part of $\FFF_{\Phi_1\wedge\neg\Phi_2}(\gamma_i)$
	is exactly $\FFF_{\Phi_1\wedge\neg\Phi_2}(\gamma_i^+)$
	while the negative part of $\FFF_{\Phi_1\wedge\neg\Phi_2}(\gamma_i)$
	is $\FFF_{\Phi_1\wedge\neg\Phi_2}(\gamma_i^-)$.
	Since $\FFF_{\Phi_1\wedge\neg\Phi_2}(\gamma_i)=\gamma_i$,
	we have $\FFF_{\Phi_1\wedge\neg\Phi_2}(\gamma_i^+)=\gamma_i^+$
	and $\FFF_{\Phi_1\wedge\neg\Phi_2}(\gamma_i^-)=\gamma_i^-$.
	So we can see that $\gamma_i^+$ and $-\gamma_i^-$ ($i\in[m]$)
	are positive solutions of~\eqref{eq:bottom}
	that together can linearly express any solution of~\eqref{eq:bottom}.
	
	Fixed a postive solution $\gamma=\sum_j \lambda_j \op{\Psi_j}{\Psi_j}$
	in the solution set
	$\{\gamma_i^+:i\in[m]\} \cup \{-\gamma_i^-:i\in[m]\} \setminus \{0\}$,
	we have
	\[
	\begin{aligned}
	\gamma-\lambda_j\FFF_{\Phi_1\wedge\neg\Phi_2}(\op{\Psi_j}{\Psi_j}) 
	& =\FFF_{\Phi_1\wedge\neg\Phi_2}(\gamma)
	-\lambda_j\FFF_{\Phi_1\wedge\neg\Phi_2}(\op{\Psi_j}{\Psi_j}) \\
	& = \FFF_{\Phi_1\wedge\neg\Phi_2}(\gamma-\lambda_j\op{\Psi_j}{\Psi_j})
	\end{aligned}
	\]
	is positive for each $\ket{\Psi_j}$ ($j\in[m]$),
	which implies
	$\supp(\FFF_{\Phi_1\wedge\neg\Phi_2}(\op{\Psi_j}{\Psi_j}))$
	is contained in $\supp(\gamma)$.
	In other words,
	for any Kraus operator $\FF_\ell$ of $\FFF_{\Phi_1\wedge\neg\Phi_2}$,
	$\FF_\ell \ket{\Psi_j}$ is in $\supp(\gamma)$,
	i.e. $(\sum_l \op{\Psi_l}{\Psi_l}) \FF_\ell \ket{\Psi_j}=\FF_\ell \ket{\Psi_j}$.
	Furthermore, for any $\ket{\Psi}\in\supp(\gamma)$,
	after expressing it as $\sum_j c_j \ket{\Psi_j}$
	with $\sum_{i\in[d]} |c_i|^2=1$, we have
	\[
	\begin{aligned}
	& \left(\sum_j \op{\Psi_j}{\Psi_j}\right)
	\FFF_{\Phi_1\wedge\neg\Phi_2}(\op{\Psi}{\Psi})
	\left(\sum_j \op{\Psi_j}{\Psi_j}\right) \\
	=\ & \left(\sum_j \op{\Psi_j}{\Psi_j}\right)
	\left[\sum_\ell \FF_\ell
	\left(\sum_j\sum_l c_j c_l^* \op{\Psi_j}{\Psi_l} \right) \FF_\ell^\dag\right]
	\left(\sum_j \op{\Psi_j}{\Psi_j}\right) \\
	=\ & \sum_\ell \FF_\ell
	\left(\sum_j\sum_l c_j c_l^* \op{\Psi_j}{\Psi_l} \right) \FF_\ell^\dag \\
	=\ & \FFF_{\Phi_1\wedge\neg\Phi_2}(\op{\Psi}{\Psi}),
	\end{aligned}
	\]
	which implies $\supp(\FFF_{\Phi_1\wedge\neg\Phi_2}(\op{\Psi}{\Psi}))$
	is contained in $\supp(\gamma)$.
	Thus $\supp(\gamma)$ is bottom w.r.t. $\FFF_{\Phi_1\wedge\neg\Phi_2}$.
	Additionally,
	$\spn(\bigcup_{k=0}^\infty \supp(\FFF^k(\op{\Psi_j}{\Psi_j})))$
	forms a BSCC subspace w.r.t. $\FFF_{\Phi_1\wedge\neg\Phi_2}$.
	Hence,
	$\supp(\gamma_i)$ is the direct-sum of some BSCC subspaces of $\h_\cq$,
	as well as $\Gamma$.
	The latter covers all fixed-points of $\FFF_{\Phi_1\wedge\neg\Phi_2}$,
	since any fixed-point of $\FFF_{\Phi_1\wedge\neg\Phi_2}$
	can be linearly expressed
	by $\{\gamma_i^+:i\in[m]\} \cup \{-\gamma_i^-:i\in[m]\}$,
	whose supports are contained in $\Gamma$.
	
	We proceed to prove that $\Gamma$ is the direct-sum of all BSCC subspaces.
	By the decomposition~\cite[Thm.~5]{YFY+13} and \cite[Thm.~1]{GFY18},
	we have
	\[
	\h_\cq=\mathcal{T} \oplus \bigoplus_{i} \Gamma_i,
	\]
	where $\mathcal{T}$ is the maximal transient subspace
	w.r.t. $\FFF_{\Phi_1\wedge\neg\Phi_2}$
	and each $\Gamma_i$ is a BSCC subspace;
	and although the decomposition is not unique,
	the maximal transient subspace $\mathcal{T}$ is unique
	as well as the direct-sum of all BSCC subspaces $\Gamma_i$.
	We assume by contradiction that
	$\Gamma$ does not contain all BSCC subspaces.
	Then there is a BSCC subspace $\Gamma_0$ orthogonal to $\Gamma$.
	It is easy to see that
	\begin{itemize}
		\item the set $\DDD_{\Gamma_0}^1$ of density operators $\rho$
		on $\Gamma_0$ with trace $1$
		is a convex and compact set
		in the viewpoint of probabilistic ensemble form
		$\{(p_i,\ket{\psi_i}):i\in[d]\}$
		that is obained from the spectral decomposition
		$\rho=\sum_{i\in[d]} p_i \op{\psi_i}{\psi_i}$; and
		\item $\FFF_{\Phi_1\wedge\neg\Phi_2}$ is a continuous function
		mapping $\DDD_{\Gamma_0}^1$ to itself.
	\end{itemize}
	By Brouwer's fixed-point theorem~\cite[Chap.~4]{Ist01} that
	for a continuous function $f$ mapping a convex and compact set $\chi$ to itself,
	there is a point $x\in\chi$ such that $f(x)=x$,
	we know there is
	a fixed-point $\rho_0$ of $\FFF_{\Phi_1\wedge\neg\Phi_2}$ in $\DDD_{\Gamma_0}^1$.
	From the construction of $\Gamma$, however,
	we have $\supp(\rho_0)\subseteq\Gamma$,
	which implies $\Gamma_0$ is not orthogonal to $\Gamma$
	and thus contradicts the assumption.
	Hence we obtain that
	$\Gamma$ is exactly
	the direct-sum of all BSCC subspaces w.r.t. $\FFF_{\Phi_1\wedge\neg\Phi_2}$. \qed
\end{proof}

We formally describe the procedure
to compute the direct-sum $\Gamma$ of all BSCC subspaces
w.r.t. $\FFF_{\Phi_1\wedge\neg\Phi_2}$
as Algorithm~\ref{alg:BSCC}.
By invoking it on the super-operator $\FFF_{\Phi_1\wedge\neg\Phi_2}$
and the Hilbert space $\h_\cq$,
we would obtain the direct-sum $\Gamma$ in $\BigO(N^6)$ \emph{arithmetic operations},
which is more efficient than the existing method~\cite[Procedure GetBSCC]{FHT+17}
in $\BigO(N^7)$ \emph{field} operations.\footnote{%
In~\cite{FHT+17}, the authors need to determine all individual BSCC subspaces,
collect those individual BSCC subspaces of the desired parity,
and thus check the $\omega$-regular properties.
To this end, \cite[Procedure GetBSCC]{FHT+17}
first compute the direct-sum of BSCC subspaces corresponding to positive eigenvalues
and the direct-sum of BSCC subspaces corresponding to negative eigenvalues.
If those direct-sums consist of more than one BSCC subspace,
the procedure would be respectively applied to the two direct-sums
in a recursive manner.
The overall complexity is $\BigO(N^7)$.
In our setting,
it suffices to compute the direct-sum of all BSCC subspaces,
which saves the recursion to complexity $\BigO(N^6)$.
Additionally, determining positive/negative eigenvalues
is a typical kind of field operations beyond arithmetic ones 
(addition, subtraction, multiplication, and division).
Obviously, the latters are of lower computational cost.}

\begin{algorithm}[H]
	\caption{\textsf{Computing the direct-sum of all BSCC subspaces.}}\label{alg:BSCC}
	\begin{algorithmic}[1]
		\item[] $$\Gamma \gets {\sf BSCC}(\EEE,\h)$$
		\Require $\EEE\in\SSS$ is a super-operator
		on the Hilbert space $\h$ of dimension $d$.
		\Ensure $\Gamma$ is the direct-sum of all BSCC subspaces w.r.t. $\EEE$.
		\State $\Gamma \gets\{0\}$;
		\Comment{initializing $\Gamma$ as the zero space}
		\State compute all linearly independent solutions
		$\gamma_i$ ($i\in[m]$) of $\EEE(\gamma)=\gamma$
		($\gamma=\gamma^\dag\in\LLL_\h$);
		\For{each $i\in[m]$}
		\State $\Gamma \gets \spn(\Gamma \cup \supp(\gamma_i))$;
		\EndFor
		\State \Return $\Gamma$.
		\Com $\BigO(d^6)$.
	\end{algorithmic}
\end{algorithm}
\paragraph{Complexity}
In Algorithm~\ref{alg:BSCC},
the stationary equation $\gamma=\EEE(\gamma)$
can be solved in $\BigO(d^6)$ by Guassian elimination,
whose complexity is cubic in the number $d^2$ of real variables in $\gamma$.
The support $\supp(\gamma_i)$ of an individual solution $\gamma_i$
and the extended space $\spn(\Gamma \cup \supp(\gamma_i))$
can be computed in $\BigO(d^3)$ by the Gram--Schmidt procedure,
whose complexity is cubic in the dimension $d$.
In total, they are in $\BigO(md^3) \subseteq \BigO(d^5)$,
as the number $m$ of linearly independent solutions is bounded by $d^2$.

Let $\PPP_\Gamma=\{\PP_\Gamma\}$
where $\PP_\Gamma$ is the projector onto $\Gamma$, i.e. $\PP_\Gamma(\h_\cq)=\Gamma$;
and $\PPP_{\Gamma^\perp}=\{\PP_{\Gamma^\perp}\}$
where $\Gamma^\perp$ is the orthogonal complement of $\Gamma$,
i.e. $\Gamma\oplus\Gamma^\perp=\h_\cq$.
Again, thanks to the fact the states in the QMC are of
the mixed structure $\rho=\sum_{s\in S} \op{s}{s}\otimes\rho_s$,
we have that $\PP_\Gamma$ is of
the form $\sum_{s\in S} \op{s}{s}\otimes\PP_s$
where $\PP_s$ ($s\in S$) are positive operators,
as well as $\PP_{\Gamma^\perp}=\ind_{\h_\cq}-\PP_\Gamma$.

\begin{example}\label{ex5}
	Consider the path formula
	$\phi_4=\true \,\ntl^{\le 15} (\ok\vee\error)$
	on the QMC $\QMC_1$ in Example~\ref{ex1}.
	The repeated super-operator in the SOVM $\Delta(\phi_4)$ is
	\begin{align*}
		&\ \FFF_{\neg\ok\wedge\neg\error}
		:= \FFF\circ \PPP_{\neg\ok\wedge\neg\error}
		=\FFF\circ \PPP_{\true\wedge\neg(\ok\vee\error)} \\
		= &\ \left\{ \begin{array}{l}
			\op{s_1}{s_0}\otimes \op{1,+}{1,1},
			\tfrac{4}{5}\op{s_1}{s_0}\otimes \op{1,-}{1,2},
			\tfrac{3}{5}\op{s_5}{s_0}\otimes \op{1,2}{1,2}, \\
			\op{s_5}{s_0}\otimes \op{2}{2}\otimes\ind,
			\op{s_0}{s_1}\otimes \op{1,1}{1,+},
			\tfrac{4}{5}\op{s_0}{s_1}\otimes \op{1,2}{1,-}, \\
			\tfrac{3}{5}\op{s_2}{s_1}\otimes \op{1,2}{1,-},
			\op{s_2}{s_1}\otimes \op{2}{2}\otimes\ind,
			\tfrac{12}{25}\op{s_0}{s_2}\otimes \mathbf{X}\otimes\ind, \\
			\tfrac{9}{25}\op{s_0}{s_2}\otimes \mathbf{X}\otimes\mathbf{X},
			\tfrac{16}{25}\op{s_3}{s_2}\otimes \ind\otimes\ind,
			\tfrac{12}{25}\op{s_3}{s_2}\otimes \ind\otimes\mathbf{X}, \\
			\tfrac{12}{25}\op{s_0}{s_3}\otimes \ind\otimes\mathbf{Z},
			\tfrac{12}{25}\op{s_0}{s_3}\otimes \mathbf{Z}\otimes\ind,
			\tfrac{16}{25}\op{s_4}{s_3}\otimes \ind\otimes\ind,
			\tfrac{9}{25}\op{s_4}{s_3}\otimes \mathbf{Z}\otimes\mathbf{Z}
		\end{array} \right\}.
	\end{align*}
	Solving the stationary equation
	$\FFF_{\neg\ok\wedge\neg\error}(\gamma)=\gamma$
	($\gamma=\gamma^\dag=\sum_{s\in S} \op{s}{s} \otimes \gamma_s \in\LLL_{\h_\cq}$),
	we obtain only one solution
	$\gamma_1=\op{s_0}{s_0}\otimes\op{1,1}{1,1}+\op{s_1}{s_1}\otimes\op{1,+}{1,+}$.

	The BSCC subspaces $\Gamma$
	covering all the fixed-points of $\FFF_{\neg\ok\wedge\neg\error}$
	is actually $\spn(\{\ket{s_0}\otimes\ket{1,1},\ket{s_1}\otimes\ket{1,+}\})$.
	The projection super-operator $\PPP_\Gamma=\{\PP_\Gamma\}$ onto $\Gamma$
	is given by the projector
	$\PP_\Gamma=\op{s_0}{s_0}\otimes\op{1,1}{1,1}+\op{s_1}{s_1}\otimes\op{1,+}{1,+}$;
	and the projection super-operator $\PPP_{\Gamma^\perp}=\{\PP_{\Gamma^\perp}\}$
	onto $\Gamma^\perp$ is given by $\PP_{\Gamma^\perp}=\ind_{\h_\cq}-\PP_\Gamma$.
	%\[
	%\begin{aligned}
	%	\PP_{\Gamma^\perp} =\ & \op{s_0}{s_0}\otimes\op{1,2}{1,2}
	%		+\op{s_0}{s_0}\otimes\op{2}{2}\otimes\ind
	%		+\op{s_1}{s_1}\otimes\op{1,-}{1,-} \ + \\
	%		& \op{s_1}{s_1}\otimes\op{2}{2}\otimes\ind
	%		+\sum\{|\,\op{s}{s}\otimes\ind\otimes\ind:s\in\{s_2,s_3,s_4,s_5\}\,|\},
	%\end{aligned}
	%\]
	%where $\{|\,\cdot\,|\}$ denotes a multiset.
	Thereby, the composite super-operator
	$\FFF_{\neg\ok\wedge\neg\error} \circ \PPP_{\Gamma^\perp}$
	would have no fixed-point.
\end{example}

\begin{lemma}\label{lem:invariant}
	The identity
	$\PPP_{\Phi_2} \circ (\FFF_{\Phi_1\wedge\neg\Phi_2})^i
	=\PPP_{\Phi_2} \circ (\FFF_{\Phi_1\wedge\neg\Phi_2} \circ \PPP_{\Gamma^\perp})^i$
	holds for each $i \ge 0$.
\end{lemma}
\begin{proof}
	We will prove it by induction on $i$.
	When $i=0$, the identity follows trivially.
	Assume the identity holds for $i<k$.
	We proceed to show that it holds for $i=k$.
	Let $\PPP_\Gamma=\{\PP_\Gamma\}$
	and $\PPP_{\Gamma^\perp}=\{\PP_{\Gamma^\perp}\}$.
	For any $\ket{\Psi} \in \h_\cq$, we have
	\begin{align*}
	& \PPP_{\Phi_2} \circ
	(\FFF_{\Phi_1\wedge\neg\Phi_2})^k(\op{\Psi}{\Psi}) \\
	=\ & \PPP_{\Phi_2} \circ (\FFF_{\Phi_1\wedge\neg\Phi_2})^k
	[(\PP_\Gamma+\PP_{\Gamma^\perp}) \op{\Psi}{\Psi}
	(\PP_\Gamma+\PP_{\Gamma^\perp})] \\
	=\ & \PPP_{\Phi_2} \circ (\FFF_{\Phi_1\wedge\neg\Phi_2})^k
	[\PPP_\Gamma(\op{\Psi}{\Psi})
	+\PP_\Gamma\op{\Psi}{\Psi}\PP_{\Gamma^\perp}
	+\PP_{\Gamma^\perp}\op{\Psi}{\Psi}\PP_\Gamma
	+\PPP_{\Gamma^\perp}(\op{\Psi}{\Psi})] \displaybreak[0] \\
	=\ & \PPP_{\Phi_2} \circ (\FFF_{\Phi_1\wedge\neg\Phi_2})^{k-1}
	[\FFF_{\Phi_1\wedge\neg\Phi_2}\circ\PPP_\Gamma(\op{\Psi}{\Psi})
	+\FFF_{\Phi_1\wedge\neg\Phi_2}
	(\PP_\Gamma\op{\Psi}{\Psi}\PP_{\Gamma^\perp}) \ + \\
	& \qquad \FFF_{\Phi_1\wedge\neg\Phi_2}
	(\PP_{\Gamma^\perp}\op{\Psi}{\Psi}\PP_\Gamma)
	+\FFF_{\Phi_1\wedge\neg\Phi_2}\circ\PPP_{\Gamma^\perp}(\op{\Psi}{\Psi})] \\
	=\ & \PPP_{\Phi_2} \circ (\FFF_{\Phi_1\wedge\neg\Phi_2})^{k-1}
	[\PPP_\Gamma\circ\FFF_{\Phi_1\wedge\neg\Phi_2}\circ\PPP_\Gamma(\op{\Psi}{\Psi})
	+\PP_\Gamma\FFF_{\Phi_1\wedge\neg\Phi_2}
	(\PP_\Gamma\op{\Psi}{\Psi}\PP_{\Gamma^\perp}) \ + \\
	& \qquad \FFF_{\Phi_1\wedge\neg\Phi_2}
	(\PP_{\Gamma^\perp}\op{\Psi}{\Psi}\PP_\Gamma) \PP_\Gamma
	+\FFF_{\Phi_1\wedge\neg\Phi_2}\circ\PPP_{\Gamma^\perp}(\op{\Psi}{\Psi})] \displaybreak[0] \\
	=\ & \PPP_{\Phi_2} \circ
	(\FFF_{\Phi_1\wedge\neg\Phi_2} \circ \PPP_{\Gamma^\perp})^{k-1}
	(\PPP_\Gamma\circ\FFF_{\Phi_1\wedge\neg\Phi_2}\circ\PPP_\Gamma(\op{\Psi}{\Psi}) 
	+ \PP_\Gamma\FFF_{\Phi_1\wedge\neg\Phi_2}
	(\PP_\Gamma\op{\Psi}{\Psi}\PP_{\Gamma^\perp}) \ + \\
	& \qquad \FFF_{\Phi_1\wedge\neg\Phi_2}
	(\PP_{\Gamma^\perp}\op{\Psi}{\Psi}\PP_\Gamma) \PP_\Gamma
	+ \FFF_{\Phi_1\wedge\neg\Phi_2}\circ\PPP_{\Gamma^\perp}(\op{\Psi}{\Psi}))
	\displaybreak[0] \\
	=\ & \PPP_{\Phi_2}
	\circ (\FFF_{\Phi_1\wedge\neg\Phi_2} \circ \PPP_{\Gamma^\perp})^{k-1}
	\circ \FFF_{\Phi_1\wedge\neg\Phi_2}
	\circ \PPP_{\Gamma^\perp}(\op{\Psi}{\Psi}) \\
	=\ & \PPP_{\Phi_2}
	\circ (\FFF_{\Phi_1\wedge\neg\Phi_2}
	\circ \PPP_{\Gamma^\perp})^k(\op{\Psi}{\Psi}),
	\end{align*}
	where the fourth equation follows from the facts:
	\begin{itemize}
		\item $\PP_\Gamma\ket{\Psi}$ is in $\Gamma$, and
		\item letting $\FF$ be a Kraus operator of $\FFF_{\Phi_1\wedge\neg\Phi_2}$,
		then $\FF\PP_\Gamma\ket{\Psi}$ is still in $\Gamma$;
	\end{itemize}
	and the sixth equation follows from the facts:
	\begin{itemize}
		\item for $k>1$, $\Gamma$ is orthogonal to $\Gamma^\perp$, and
		\item for $k=1$, letting $\PPP_{\Phi_1\wedge\neg\Phi_2}
		=\{\PP_{\Phi_1\wedge\neg\Phi_2}\}$
		and $\PPP_{\neg\Phi_2}=\{\PP_{\neg\Phi_2}\}$,
		then $\Gamma \subseteq \PP_{\Phi_1\wedge\neg\Phi_2}(\h_\cq)
		\subseteq \PP_{\neg\Phi_2}(\h_\cq)$
		is orthogonal to $\PP_{\Phi_2}(\h_\cq)$. \qed
	\end{itemize}
\end{proof}

We are going to represent $\Delta(\Phi_1 \ntl \Phi_2)$ using explicit matrices.
Recall from \cite[Def.~2.2]{YYF13} that,
given a super-operator $\EEE=\{\EE_\ell: \ell\in[m]\}$,
it has the matrix representation
\begin{equation}\label{eq:s2m}
\sm(\EEE):=\sum_{\ell \in [m]} \EE_\ell \otimes \EE_\ell^*,
\end{equation}
where $*$ denotes complex conjugate.
%Then for any $\gamma \in \LLL$, we have
%\begin{equation}
%$\EEE(\gamma) = \sum_{i,j \in [d]} \bra{i,j} \sm(\EEE)
%\linebreak[0](\gamma \otimes \ind) \sum_{l \in [d]} \ket{l,l} \op{i}{j}$
%\end{equation}
Let
\begin{itemize}
	\item $\mv(\gamma):=\sum_{i,j \in [n]} \bra{i}\gamma\ket{j} \ket{i,j}$
	be the function that
	rearranges entries of the linear operator $\gamma$ as a column vector;
	and
	\item $\vm(\mathbf{v}):=\sum_{i,j \in [n]} \bra{i,j} \mathbf{v} \op{i}{j}$
	be the function that
	rearranges entries of the column vector $\mathbf{v}$ as a linear operator.
\end{itemize}
Here, $\sm$, $\mv$ and $\vm$ are read as ``super-operator to matrix'',
``linear operator to vector'' and ``vector to linear operator'', respectively.
Then, we have the identities $\vm(\mv(\gamma))=\gamma$,
$\mv(\EEE(\gamma))=\sm(\EEE)\mv(\gamma)$,
and $\sm(\EEE_2 \circ \EEE_1)=\sm(\EEE_2) \sm(\EEE_1)$.
Therefore, all involved super-operator manipulations
can be converted to matrix manipulations.

Suppose that all classical states in $S$ are ordered as $s_1 \prec \cdots \prec s_n$
where $s_1$ is the initial one, i.e. $\Omega=Path(s_1)$.
We notice that for any classical-quantum state
$\rho=\sum_{i\in[n]} \op{s_i}{s_i}\otimes\rho_i$ in $\DDD_{\h_\cq}$,
$\FFF_{\Phi_1\wedge\neg\Phi_2} \circ \PPP_{\Gamma^\perp}(\rho)$
and $\PPP_{\Phi_2}(\rho)$
keep the mixed form $\sum_{i\in[n]} \op{s_i}{s_i}\otimes\rho_i'$
for some $\rho_i' \in \DDD$.
So, we can compressively define the matrix representation of $\rho$
as a column vector,
consisting of $n$ blocks as entries,
in which the $i$th entry is the column vector $\mv(\rho_i)$ for $\rho_i$,
i.e. $\M_1 = \sum_{i\in[n]} \ket{i} \otimes \mv(\rho_i)$.
%\begin{equation}
%	\M_0 = \sum_{i\in[n]} \ket{i} \otimes \mv(\rho_i)
%	= \begin{bmatrix}
%		\mv(\rho_1)	\\
%		\vdots \\
%		\mv(\rho_n)
%	\end{bmatrix}.
%\end{equation}

Let $\FFF_{\Phi_1\wedge\neg\Phi_2}
=\sum_{i,j \in [n]} \{\op{s_j}{s_i}\} \otimes Q(s_i,s_j)
=\bigcup_{i,j \in [n]} \bigcup_\ell \{\op{s_j}{s_i} \otimes \mathbf{Q}_{i,j,\ell}\}$
where $\mathbf{Q}_{i,j,\ell}$ are Kraus operators of $Q(s_i,s_j)$
and $\PPP_{\Gamma^\perp}=\{\sum_{k \in [n]} \op{s_k}{s_k} \otimes \PP_k\}$.
Then, $\FFF_{\Phi_1\wedge\neg\Phi_2} \circ \PPP_{\Gamma^\perp}$ is
\begin{equation}
	\bigcup_{i,j \in [n]} \bigcup_\ell
	\left\{\sum_{k \in [n]} \op{s_j}{s_i} \op{s_k}{s_k}
	\otimes \mathbf{Q}_{i,j,\ell} \PP_k\right\}
	= \bigcup_{i,j \in [n]} \bigcup_\ell
	\left\{\op{s_j}{s_i} \otimes \mathbf{Q}_{i,j,\ell} \PP_i\right\}.
\end{equation}
Using it, we further define:
\begin{subequations}\label{eq:matrices}
\begin{itemize}
%	\item the matrix representation of the projection super-operator $\PPP_s$
%	as a diagonal matrix,
%	consisting of $n$ blocks as diagonal entries,
%	in which the first entry is $\ind_{\h\otimes\h}$ and others are $0$,
%	i.e. $\M_1 = \op{1}{1} \otimes \ind_{\h\otimes\h}$;
	\item the matrix representation of
	$\FFF_{\Phi_1\wedge\neg\Phi_2} \circ \PPP_{\Gamma^\perp}$ as a square matrix,
	consisting of $n^2$ blocks as entries,
	in which the $(j,i)$-th entry is $\sum_\ell
	\mathbf{Q}_{i,j,\ell} \PP_i \otimes \mathbf{Q}_{i,j,\ell}^* \PP_i^*$,
	i.e.
	\begin{equation}
		\M_2 = \sum_{i,j \in [n]} \sum_\ell \op{j}{i} \otimes
		\mathbf{Q}_{i,j,\ell} \PP_i	\otimes \mathbf{Q}_{i,j,\ell}^* \PP_i^*;
	\end{equation}
	\item the matrix representation of the projection super-operator $\PPP_{\Phi_2}$
	as a diagonal matrix,
	consisting of $n$ blocks as diagonal entries,
	in which the $i$th entry is $\ind_{\h\otimes\h}$
	if the predicate $s_i\models\Phi_2$ is true,
	and $0$ otherwise,
	i.e.
	\begin{equation}
		\M_3 =\sum \{|\, \op{i}{i} \otimes \ind_{\h\otimes\h}:
		i\in[n] \wedge s_i\models\Phi_2 \,|\},
	\end{equation}
	where $\{|\,\cdot\,|\}$ denotes a multiset.
\end{itemize}
All these matrix representations are obtained
by extending~\eqref{eq:s2m} on $\h$ to the enlarged space $\h_\cq$. 
\end{subequations}

\begin{lemma}\label{lem:invertible}
	The matrix $\ind_{\h_\cq\otimes\h}-\M_2$ is invertible.
\end{lemma}
\begin{proof}
	It suffices to show $\M_2$ has no eigenvalue $1$.
	We assume by contradiction that
	there is an eigenvector $\mathbf{v}$ of $\M_2$
	associated with eigenvalue $1$.
	That is, $\M_2 \mathbf{v}=\mathbf{v} \ne 0$. 
	Then,
	\[
	\gamma=\sum_{i\in[n]} \op{s_i}{s_i} \otimes
	\vm((\bra{i} \otimes \ind_{\h\otimes\h})\mathbf{v})
	\]
	is a linear operator on $\h_\cq$,
	satisfying
	$\FFF_{\Phi_1\wedge\neg\Phi_2} \circ \PPP_{\Gamma^\perp}(\gamma)=\gamma \ne 0$,
	while $\gamma_0=\gamma+\gamma^\dag$ also a linear operator on $\h_\cq$,
	satisfying
	$\FFF_{\Phi_1\wedge\neg\Phi_2} \circ \PPP_{\Gamma^\perp}(\gamma_0) =\gamma_0 \ne 0$.
	By the definition of $\Gamma$, we have $\supp(\gamma_0) \subseteq \Gamma$,
	and thus $\FFF_{\Phi_1\wedge\neg\Phi_2} \circ \PPP_{\Gamma^\perp}(\gamma_0)
	=\FFF_{\Phi_1\wedge\neg\Phi_2}(0)=0 \ne \gamma_0$,
	which contradicts the assumption. \qed
\end{proof}

\begin{theorem}[Matrix representation]\label{thm:matrix}
	Let $\M_2$ and $\M_3$ be the matrices as defined in~\eqref{eq:matrices}.
	Then it is in polynomial time to obtain:
    \begin{enumerate}
        \item the explicit matrix representation of
        the super-operator $\Delta(\nxt \Phi)$ as
		$\sum_{s\models\Phi} \sm(Q(s_1,s))$,
		\item the explicit matrix representation of
	    $\Delta(\Phi_1 \ntl^{\le k} \Phi_2)$ as
	    \[
			\sum_{i\in[n]} (\bra{i} \otimes \ind_{\h\otimes\h})
			\M_3 (\ind_{\h_\cq\otimes\h}-\M_2^{k+1})
			(\ind_{\h_\cq\otimes\h}-\M_2)^{-1}
			(\ket{1} \otimes \ind_{\h\otimes\h}),
	    \]
        \item the explicit matrix representation of
        $\Delta(\Phi_1 \ntl \Phi_2)$ as  
	    \[
			\sum_{i\in[n]} (\bra{i} \otimes \ind_{\h\otimes\h})
			\M_3 (\ind_{\h_\cq\otimes\h}-\M_2)^{-1}
			(\ket{1} \otimes \ind_{\h\otimes\h}).
        \]
    \end{enumerate}
\end{theorem}
\begin{proof}
	The matrix representations directly follow from
	the semantics of the next formula $\nxt \Phi$,
	the bounded-until formula $\Delta(\Phi_1 \ntl^{\le k} \Phi_2)$,
	and the unbounded-until formula $\Delta(\Phi_1 \ntl \Phi_2)$.
	For complexity, we will analyze them in turn.
	
	\begin{enumerate}
		\item It is a sum of at most $nd^2$ matrix tensor products,
		each costs $\BigO(d^4)$.
		In total, it is in $\BigO(nd^6) \subseteq \BigO(N^6)$.
		\item The matrix $\ind_{\h_\cq\otimes\h}-\M_2$ is of dimension $nd^2$.
		Computing its inverse costs $\BigO(n^3d^6)$.
		The matrix power $\M_2^{k+1}$ amounts to
		\[
		\M_2^{b_0 \cdot 2^0} \M_2^{b_1 \cdot 2^1} \cdots \M_2^{b_l \cdot 2^l},
		\]
		where $(b_l,\ldots,b_1,b_0)$ is the binary code of the positive integer $k+1$
		with $l=\lceil\log_2(k+2)\rceil-1$,
		i.e. $k+1=b_0 \cdot 2^0+b_1 \cdot 2^1 \cdots b_l \cdot 2^l$
		with $b_j \in\{0,1\}$.
		Computing $\M_2^{k+1}$ requires sequentially computing
		all the factors $\M_2^{2^j}$ ($j\in[l]$),
		each of which costs $\BigO(n^3d^6)$;
		and then computing the product of those factors corresponding to $b_j=1$,
		which costs $\BigO(n^3d^6\log_2(k))$.
		Other operations are merely a few matrix-vector multiplications
		over an $nd^2$-dimensional vector space,
		which costs $\BigO(n^2d^4)$.
		Totally, it is in $\BigO(n^3d^6\log_2(k)) \subseteq \BigO(N^6\log_2(k))$.
		\item It is clearly in $\BigO(N^6)$ by the previous analysis.
	\end{enumerate}
	
	As a result, the complexity is polynomial time
	w.r.t. $N=nd$ (reflected in the size of $\QMC$)
	and linear time w.r.t. $\log_2(k)$ (reflected in the size of $\phi$). \qed
\end{proof}

\begin{example}\label{ex6}
	Consider the path formulas
	% $\phi_4=\true\,\ntl^{\le 15} (\ok\vee\error)$
	\[
		\begin{aligned}
		& \phi_1=\true\,\ntl \ok,
		&& \phi_2=\true\,\ntl^{\le 15} \ok, \\
		& \phi_3=\true\,\ntl (\ok\vee\error),
		&& \phi_4=\true\,\ntl^{\le 15} (\ok\vee\error)
		\end{aligned}
	\]
	on the QMC $\QMC_1$ shown in Example~\ref{ex1}.
	For $\phi_4$, the repeated super-operator $\FFF_{\neg\ok \wedge \neg\error}$
	and the projector $\PP_\Gamma$ whose support covers all its fixed-points
	have been computed in Example~\ref{ex5}.
	%and that the composite super-operator
	%$\FFF_{\neg\ok \wedge \neg\error} \circ \PPP_{\Gamma^\perp}$ is
	%\[
	%	\begin{aligned}
	%	\{ & \tfrac{4}{5}\op{s_1}{s_0}\otimes \op{1,-}{1,2},
	%	\tfrac{3}{5}\op{s_5}{s_0}\otimes \op{1,2}{1,2},
	%	\op{s_5}{s_0}\otimes \op{2}{2}\otimes\ind, \\
	%	& \tfrac{4}{5}\op{s_0}{s_1}\otimes \op{1,2}{1,-},
	%	\tfrac{3}{5}\op{s_2}{s_1}\otimes \op{1,2}{1,-},
	%	\op{s_2}{s_1}\otimes \op{2}{2}\otimes\ind, \\
	%	& \tfrac{12}{25}\op{s_0}{s_2}\otimes \mathbf{X}\otimes\ind,
	%	\tfrac{9}{25}\op{s_0}{s_2}\otimes \mathbf{X}\otimes\mathbf{X},
	%	\tfrac{16}{25}\op{s_3}{s_2}\otimes \ind\otimes\ind,
	%	\tfrac{12}{25}\op{s_3}{s_2}\otimes \ind\otimes\mathbf{X}, \\
	%	& \tfrac{12}{25}\op{s_0}{s_3}\otimes \ind\otimes\mathbf{Z},
	%	\tfrac{12}{25}\op{s_0}{s_3}\otimes \mathbf{Z}\otimes\ind,
	%	\tfrac{16}{25}\op{s_4}{s_3}\otimes \ind\otimes\ind,
	%	\tfrac{9}{25}\op{s_4}{s_3}\otimes \mathbf{Z}\otimes\mathbf{Z} \}.
	%	\end{aligned}
	%\]
	Under the order $s_0 \prec \cdots \prec s_5$,
	the matrix representations are calculated as
	\begin{align*}
%	\M_1 =\ & \op{4}{4} \otimes \ind_{\h\otimes\h}, \\ 
	\M_2 =\
	& \tfrac{16}{25} \op{2}{1}\otimes \op{1,-}{1,2}\otimes\op{1,-}{1,2}
	+ \tfrac{9}{25} \op{6}{1}\otimes \op{1,2}{1,2}\otimes \op{1,2}{1,2} + \\
	& \op{6}{1}\otimes \op{2}{2}\otimes\ind\otimes \op{2}{2}\otimes\ind
	+ \tfrac{16}{25} \op{1}{2}\otimes \op{1,2}{1,-}\otimes \op{1,2}{1,-} \ + \\
	& \tfrac{9}{25} \op{3}{2}\otimes \op{1,2}{1,-}\otimes \op{1,2}{1,-}
	+ \op{3}{2}\otimes \op{2}{2}\otimes\ind\otimes \op{2}{2}\otimes\ind \ +
	\displaybreak[0] \\
	& \tfrac{144}{625} \op{1}{3}\otimes \mathbf{X}\otimes\ind\otimes
	\mathbf{X}\otimes\ind +
	\tfrac{81}{625} \op{1}{3}\otimes \mathbf{X}\otimes\mathbf{X}\otimes
	\mathbf{X}\otimes\mathbf{X} +
	\tfrac{256}{625} \op{4}{3}\otimes \ind\otimes\ind\otimes \ind\otimes\ind \ + \\
	& \tfrac{144}{625} \op{4}{3}\otimes \ind\otimes\mathbf{X}\otimes
	\ind\otimes\mathbf{X}
	+ \tfrac{144}{625} \op{1}{4}\otimes \ind\otimes\mathbf{Z}\otimes
	\ind\otimes\mathbf{Z} +
	\tfrac{144}{625} \op{1}{4}\otimes \mathbf{Z}\otimes\ind\otimes
	\mathbf{Z}\otimes\ind \ + \\
	& \tfrac{256}{625} \op{5}{4}\otimes \ind\otimes\ind\otimes \ind\otimes\ind + 
	\tfrac{81}{625} \op{5}{4}\otimes \mathbf{Z}\otimes\mathbf{Z}\otimes
	\mathbf{Z}\otimes\mathbf{Z}, \\		
	\M_3 =\ & \op{5}{5} \otimes \ind\otimes\ind +\op{6}{6} \otimes \ind\otimes\ind,
	\end{align*}
	in which all eigenvalues of $\M_2$ are $\pm\tfrac{8}{125}\sqrt{50+2\sqrt{1273}}$,
	$\pm\tfrac{8\mathrm{i}}{125}\sqrt{50+2\sqrt{1273}}$
	and $0$ of multiplicity $92$.
	Since $\M_2$ has no eigenvalue $1$,
	the matrix inverse $(\ind_{\h_\cq\otimes\h}-\M_2)^{-1}$ is well-defined as expected.
	Finally,
	the explicit matrix representation $\sm(\Delta(\phi_4))$ of $\Delta(\phi_4)$
	is obtained as
	\[
	\begin{aligned}
	&\sum_{i\in[6]} (\bra{i} \otimes \ind_{\h\otimes\h})
	\M_3 (\ind_{\h_\cq\otimes\h}-\M_2^{16})
	(\ind_{\h_\cq\otimes\h}-\M_2)^{-1}
	(\ket{4} \otimes \ind_{\h\otimes\h})\\
	=\ &\tfrac{7}{25}\ind\otimes\ind\otimes\ind\otimes\ind
	+\tfrac{135172248480317003605337382912}{5684341886080801486968994140625}
	\op{1,1}{1,2}\otimes\op{1,1}{1,2} \ + \\
	&\tfrac{1952505842866906373900886182688}{5684341886080801486968994140625}
	\op{1,2}{1,2}\otimes\op{1,2}{1,2} \ + \\
	&\tfrac{162}{625}(\op{1,2}{1,2}+\op{2,1}{2,1})\otimes
	(\op{1,2}{1,2}+\op{2,1}{2,1}) \ + \\
	&\tfrac{162}{625}(\op{1,1}{1,1}+\op{2,2}{2,2})\otimes
	(\op{1,1}{1,1}+\op{2,2}{2,2})
	+\tfrac{288}{625}\op{2,1}{2,1}\otimes\op{2,1}{2,1} \ + \\
	&\tfrac{288}{625}\op{2,2}{2,2}\otimes\op{2,2}{2,2}
	+\tfrac{225621334629609241922855424}{9094947017729282379150390625}\op{2,1}{1,2}
	\otimes\op{2,1}{1,2} \ + \\
	&\tfrac{401104594897083096751742976}{9094947017729282379150390625}\op{2,2}{1,2}
	\otimes\op{2,2}{1,2}.
	\end{aligned}
	\]
	Similarly, we get other matrix representations
	\begin{align*}
	\sm(\Delta(\phi_1))=\
	&\tfrac{4500000}{14835977} \op{1,2}{1,2}\otimes\op{1,2}{1,2}
	+\tfrac{288}{625} \op{2,1}{2,1}\otimes\op{2,1}{2,1} \ + \\
	&\tfrac{288}{625} \op{2,2}{2,2}\otimes\op{2,2}{2,2}
	+\tfrac{373248}{14835977}\op{2,1}{1,2}\otimes\op{2,1}{1,2} \ + \\
	&\tfrac{663552}{14835977}\op{2,2}{1,2}\otimes\op{2,2}{1,2}; \displaybreak[0] \\
	\sm(\Delta(\phi_2))=\
	&\tfrac{68487984933853712477433677856}{227373675443232059478759765625}
	\op{1,2}{1,2}\otimes\op{1,2}{1,2} \ + \\
	&\tfrac{288}{625}\op{2,1}{2,1}\otimes\op{2,1}{2,1}+\tfrac{288}{625}\op{2,2}{2,2}
	\otimes\op{2,2}{2,2} \ + \\
	&\tfrac{225621334629609241922855424}{9094947017729282379150390625}\op{2,1}{1,2}
	\otimes\op{2,1}{1,2} \ +\\
	&\tfrac{401104594897083096751742976}{9094947017729282379150390625}\op{2,2}{1,2}
	\otimes\op{2,2}{1,2}; \displaybreak[0] \\
	\sm(\Delta(\phi_3))=\
	&\tfrac{7}{25}\ind\otimes\ind\otimes\ind\otimes\ind
	+\tfrac{223617024}{9272485625}
	\op{1,1}{1,2}\otimes\op{1,1}{1,2} \ + \\
	&\tfrac{3210041376}{9272485625}
	\op{1,2}{1,2}\otimes\op{1,2}{1,2} \ + \\
	&\tfrac{162}{625}(\op{1,2}{1,2}+\op{2,1}{2,1})\otimes
	(\op{1,2}{1,2}+\op{2,1}{2,1}) \ + \\
	&\tfrac{162}{625}(\op{1,1}{1,1}+\op{2,2}{2,2})\otimes
	(\op{1,1}{1,1}+\op{2,2}{2,2}) \ + \\
	&\tfrac{288}{625}\op{2,1}{2,1}\otimes\op{2,1}{2,1}+\tfrac{288}{625}\op{2,2}{2,2}
	\otimes\op{2,2}{2,2} \ + \\
	&\tfrac{373248}{14835977}\op{2,1}{1,2}\otimes\op{2,1}{1,2} 
	+\tfrac{663552}{14835977}\op{2,2}{1,2}\otimes\op{2,2}{1,2}.  %\tag*{\qed}
	\end{align*}
\end{example}

%\begin{remark}
%	In fact, the matrix representations of the SOVM $\Delta(\phi)$
%	have the ability of quantum state tomograph.
%	Utilizing it, we can further determine
%	the explicit Kraus representations of those SOVM $\Delta(\phi)$
%	by quantum process tomograph~\cite[Sect.~8.4.2]{NiC00},
%	which costs additionally $\BigO(N^{12})$ or more.
%	So, to be more efficient, we will use the matrix representations afterwards.
%\end{remark}

\subsection{Deciding the fidelity-quantifier formula}
In the previous subsection,
we have constructed an explicit matrix representation $\M:=\sm(\EEE)$
for $\EEE=\Delta(\phi)$ where $\phi$ is the path formula
in the fidelity-quantifier formula $\QQ_{\sim\tau}(\phi)$.
Now we present an algebraic approach
to compare the (minimum) fidelity $\mfid(\EEE)$ with the threshold $\tau$,
so that $s\models\QQ_{\sim\tau}(\phi)$ can be decided.

We first notice that:
\begin{itemize}
\begin{subequations}\label{eq:quantifieds}
	\item $s\models\QQ_{\le\tau}(\phi)$ amounts to the quantified constraint
\begin{align}
	\zeta_1 & \equiv \exists\,\ket{\psi} \in \h:
		[\fid(\EEE,\op{\psi}{\psi}) \le \tau \wedge
		\forall\,\ket{\varphi} \in \h:
		\fid(\EEE,\op{\psi}{\psi}) \le \fid(\EEE,\op{\varphi}{\varphi})] \notag \\
	&\label{eq:quantified_1} \equiv
	\exists\,\ket{\psi} \in \h: \fid(\EEE,\op{\psi}{\psi}) \le \tau;
\end{align}
	\item $s\models\QQ_{\ge\tau}(\phi)$ amounts to the quantified constraint
\begin{align}
	\zeta_2 & \equiv \exists\,\ket{\psi} \in \h:
		[\fid(\EEE,\op{\psi}{\psi}) \ge \tau \wedge
		\forall\,\ket{\varphi} \in \h:
		\fid(\EEE,\op{\psi}{\psi}) \le \fid(\EEE,\op{\varphi}{\varphi})] \notag \\
	&\label{eq:quantified_2} \equiv
	\forall\,\ket{\psi} \in \h: \fid(\EEE,\op{\psi}{\psi}) \ge \tau;
\end{align}
	\item other comparison operators $=$, $<$, $>$ and $\ne$
	can be easily derived by logic connectives.
	%\item $s\models\QQ_{=\tau}(\phi)$ amounts to $\zeta_1 \wedge \zeta_2$;
	%\item $s\models\QQ_{<\tau}(\phi)$ amounts to $\neg\zeta_2$;
	%\item $s\models\QQ_{>\tau}(\phi)$ amounts to $\neg\zeta_1$; and
	%\item $s\models\QQ_{\ne\tau}(\phi)$ amounts to $\neg\zeta_1 \vee \neg\zeta_2$.
\end{subequations}
\end{itemize}

Suppose all entries in the Kraus operators $\EE$ of $\EEE$ are algebraic numbers
for the consideration of computability.
Recall that:
\begin{definition}
	A number $\lambda$ is \emph{algebraic},
	denoted by $\lambda \in \mathbb{A}$,
	if there is a nonzero $\mathbb{Q}$-polynomial $f(z)$ of least degree,
	satisfying $f(\lambda)=0$.
	Such a polynomial $f(z)$ is called
	the \emph{minimal polynomial} $f_\lambda$ of $\lambda$.
\end{definition}

Clearly, algebraic numbers widely occur in quantum information,
such as the definition of the most common quantum state
$\ket{\pm}=(\ket{1}\pm\ket{2})/\sqrt{2}$.
We will formulate the constraints~\eqref{eq:quantifieds}
as $\mathbb{Q}$-polynomial (polynomial with rational coefficients) formulas 
in the decidable theory---real closed fields~\cite{Tar51}:
% that are composed
%from $\mathbb{Q}$-polynomial equations and inequalities (as atomic formulas)
%using logic connectives
%``$\neg,\wedge,\vee,\rightarrow,\leftrightarrow$''
%and quantifiers ``$\forall,\exists$''.
%In particular, the resulting formulas would be sentences in prenex normal form
%with variables being all existentially or all universally quantified.

\begin{definition}
The theory of \emph{real closed fields} is
a first-order theory $Th(\mathbb{R};+,\,\cdot\,;\linebreak[0] =,>;0,1)$,
in which
\begin{itemize}
\item the domain is $\mathbb{R}$,
\item the functions are addition `$+$' and multiplication `$\cdot$',
\item the predicates are equality `$=$' and order `$>$', and
\item the constants are $0$ and $1$.
\end{itemize}
\end{definition}

Roughly speaking,
the elements in $Th(\mathbb{R};+,\,\cdot\,;=,>;0,1)$
are $\mathbb{Q}$-polynomial formulas that are composed
from polynomial equations and inequalities (as atomic formulas)
using logic connectives
``$\neg,\wedge,\vee,$
$\rightarrow,\leftrightarrow$''
and quantifiers ``$\forall,\exists$''.

The constraints~\eqref{eq:quantifieds} are the \emph{sentences}---%
the formulas whose variables $\ket{\psi}$ are all existentially/universally quantified,
i.e. no free variable. 
We will encode them as $\mathbb{A}$-polynomial formulas,
and further encode them as $\mathbb{Q}$-polynomial formulas.

Since $\op{\psi}{\psi}$ is pure, 
we predefine $\ket{\psi}=\sum_{i\in [d]}x_i \ket{i}$
where $x_i$ ($i\in[d]$) are complex parameters, 
subject to $\sum_{i\in[d]} x_ix_i^*=1$. 
Under the purity, we have 
\begin{align}
	\fid(\EEE,\op{\psi}{\psi}) \le \tau \
	& \equiv \ \bra{\psi} \EEE(\op{\psi}{\psi}) \ket{\psi} \le \tau^2 \notag \\ 
	& \equiv\  \left(\sum_{i\in [d]}x_i^* \bra{i} \right) 
	\EEE \left(\sum_{i,j \in [d]}x_i x_j^* \op{i}{j}\right) 
	\left(\sum_{j\in [d]}x_j \ket{j}\right) \le \tau^2 \notag \\
	&\label{eq:comparison} \equiv\ 
	\left(\sum_{i,j\in [d]}x_i^*x_j \bra{i,j} \right) 
	\M \left(\sum_{i,j \in [d]}x_i x_j^* \ket{i,j}\right) \le \tau^2,
\end{align}
which results in an $\mathbb{A}$-polynomial formula.
Denote all parameters introduced here by $\x=(x_i)_{i\in [d]}$.
Further, we encode the constraint~\eqref{eq:quantified_1} as 
\begin{subequations}\label{eq:polynomials}
\begin{equation}\label{eq:polynomial_1}
	\zeta_1 \equiv \exists\,\x: \left[
		\sum_{i\in [d]}x_i x_i^*=1 \wedge
		\left(\sum_{i,j\in [d]}x_i^*x_j \bra{i,j} \right)
		\M \left(\sum_{i,j \in [d]}x_i x_j^* \ket{i,j}\right)
		\le \tau^2 \right],
\end{equation}
which is the desired $\mathbb{A}$-polynomial formula, 
involving at most 
\begin{itemize}
	\item $2d$ real variables (converted from $d$ complex variables $\x$)
	for expressing $\ket{\psi}$, 
	\item one quadratic equation for the purity, and 
	\item one quartic inequality for the comparison.
\end{itemize}

Similarly, the $\mathbb{A}$-polynomial formula
for encoding the constraint~\eqref{eq:quantified_2} is 
\begin{equation}\label{eq:polynomial_2}
	\zeta_2 \equiv \forall\,\x: \left[
		\sum_{i\in [d]}x_i x_i^*=1 \rightarrow
		\left(\sum_{i,j\in [d]}x_i^*x_j \bra{i,j} \right)
		\M \left(\sum_{i,j \in [d]}x_i x_j^* \ket{i,j}\right)
		\ge \tau^2	\right].
\end{equation}
\end{subequations}

Suppose the input $\EEE$ involves
real algebraic numbers $\Lambda=\{\lambda_j:j\in[e]\}$.
Then the $\mathbb{A}$-polynomial formulas~\eqref{eq:polynomials}
are named by $\zeta_1(\Lambda)$ and $\zeta_2(\Lambda)$, respectively.
To effectively tackle them,
we resort to the standard encoding of real algebraic number $\lambda$
that uses minimal polynomial $f_\lambda$
plus isolation interval $I_\lambda$,
which is given by linear inequalities,
like $z \in I_\lambda \equiv L<z<U$
for some rational endpoints $L$ and $U$ of $I_\lambda$,
to distinguish $\lambda$ from other real roots of $f_\lambda$.
In such a way, to encode each real algebraic number $\lambda$,
we introduce at most
\begin{itemize}
\item one real variable $z$,
\item one equation $f_\lambda=0$ of degree $\deg(f_\lambda)$, and
\item two linear inequalities $z>L$ and $z<U$
    from the isolation interval $I_\lambda$ of $\lambda$.
\end{itemize}
For instance, the aforementioned algebraic number $1/\sqrt{2}$ occurring in $\ket{\pm}$
can be encoded as the unique solution to $z^2=\tfrac{1}{2} \wedge 0<z<1$.

The $\mathbb{A}$-polynomial formulas
$\zeta_1(\Lambda)$ and $\zeta_2(\Lambda)$
can be rewritten as the $\mathbb{Q}$-polynomial ones:
\begin{subequations}\label{eq:polynomials_1}
\begin{align}
	\label{eq:polynomial_3}
	\zeta_1(\Lambda) &\equiv \exists\,\z: \left[\bigwedge_{j\in[e]}
	(f_{\lambda_j}(z_j)=0 \wedge z_j \in I_{\lambda_j})
	\wedge \zeta_1(\z) \right] \\
	\label{eq:polynomial_4}
	\zeta_2(\Lambda) &\equiv \forall\,\z: \left[\bigwedge_{j\in[e]}
	(f_{\lambda_j}(z_j)=0 \wedge z_j \in I_{\lambda_j})
	\rightarrow \zeta_2(\z) \right],
\end{align}
\end{subequations}
where $\z=(z_j)_{j\in[e]}$ are real variables
introduced to symbolize $\Lambda$.
Note that the existential quantifier $\exists\,\z$
and the universal quantifier $\forall\,\z$ can be mutually converted here,
since for each $j\in[e]$,
there is a unique solution (i.e. $\lambda_j$) to
the subformula $f_{\lambda_j}(z_j)=0 \wedge z_j \in I_{\lambda_j}$
by the standard encoding of $\lambda_j$.

Finally, applying the existential theory of the reals~\cite[Thm.~13.13]{BPR06},
we obtain:
\begin{theorem}[Decidability]\label{thm:decide}
	It is in exponential time
	to decide the fidelity-quantifier formula $\QQ_{\sim\tau}(\phi)$.
\end{theorem}
\begin{proof}
	It suffices to show that
	the formulating subprocedure is in polynomial time,
	and that the deciding subprocedure is in exponential time. 
	
	The encoding on the purity is plainly in $\BigO(d)$.
	Encoding the left hand side of the comparison
	(e.g. the formula~\eqref{eq:comparison}) involves a few matrix-vector multiplications
	over a $d^2$-dimensional vector space,
	which costs $\BigO(d^4)$.
	Thus encoding the polynomial formulas~\eqref{eq:polynomials_1}
	is in $\BigO(d^4)$,
	which means that the formulating subprocedure is in polynomial time.
	
	Then we tackle the deciding subprocedure,
	which invokes the following Algorithm~\ref{alg:QE}
	on the formulas~\eqref{eq:polynomials_1}. 
	Technically, the formulas~\eqref{eq:polynomials_1} have
	\begin{itemize}
		\item a block of $2d+e$ real variables $\x$ and $\z$
		quantified all by `$\exists$' for~\eqref{eq:polynomial_3}
		or all by `$\forall$' for~\eqref{eq:polynomial_4}, and
		\item at most $C=2+3e$ distinct polynomials
		of degree at most $D=\max(4,\linebreak[0]\max_{j\in[e]}\deg(f_{\lambda_j}))$.
	\end{itemize}
	Thereby, the complexity is in $C^{2d+e+1}D^{\BigO(2d+e)}$,
	an exponential hierarchy. \qed
\end{proof}

\begin{algorithm}[H]
	\caption{\textsf{Existential Theory of the Reals}~\cite[Thm.~13.13]{BPR06}.}\label{alg:QE}
	\begin{algorithmic}[1]
		\item[] $$\true/\false \gets {\sf QE}(\Q\,\x: F(\x))$$
		\Require $\Q\,\x: F(\x)$ is a quantified polynomial formula, in which
		\begin{itemize}
			\item $\x$ is a block of $k$ real variables,
			which is quantified by $\Q \in \{\forall,\exists\}$,
			\item each atomic formula in $F$ is in the form $p \sim 0$
			where $\sim\,\in \{<,\le,=,\ge,>,\ne\}$,
			\item all distinct polynomials $p$,
			regardless of a constant factor,
			extracted from those atomic formulas $p \sim 0$
			form a polynomial collection $\mathbb{P}$,
			\item $C$ is the cardinality of $\mathbb{P}$, and
			\item $D$ is the maximum degree of the polynomials in $\mathbb{P}$.
		\end{itemize}
		\Ensure $\true/\false$ is the truth of $\Q\,\x: F(\x)$.
		\Com $C^{k+1} D^{\BigO(k)}$.
	\end{algorithmic}
\end{algorithm}

There are many packages that have implemented Algorithm~\ref{alg:QE},
such as \textsc{Reduce} (a.k.a. \textsc{Redlog}~\cite{DoS97})
and \textsc{Z3}~\cite{dMB08}.

\begin{example}	
	Consider the events that
	%the IP address is properly or wrongly configured within 15 steps
	\begin{enumerate}
	\item ``\textit{the IP address is properly configured}'',
	\item ``\textit{the IP address is properly configured within 15 steps}'',
	\item ``\textit{the IP address is properly or wrongly configured}'', and
	\item ``\textit{the IP address is properly or wrongly configured within 15 steps}''
	\end{enumerate}
	on the QMC $\QMC_1$ shown in Example~\ref{ex1},
	which are specified by
	the path formulas $\phi_1$ through $\phi_4$ in Example~\ref{ex6}, respectively.
	For $\phi_4=\true\, \ntl^{\le 15} (\ok\vee\error)$,
	the explicit matrix representation $\M$ of $\Delta(\phi_4)$ has been obtained.
	Now we are to decide the fidelity-quantifier formula $\QQ_{\le\tau}(\phi_4)$.

	After introducing the real variables
	$\boldsymbol{\upmu}=\Re(\mathbf{x})$ and $\boldsymbol{\upnu}=\Im(\mathbf{x})$
	where $\x=(x_i)_{i\in[4]}$ encodes the pure state $\op{\psi}{\psi}$,
	we have the desired polynomial formula 
	
\[
	\begin{aligned}
	& \exists\,\{\boldsymbol{\upmu},\boldsymbol{\upnu}\}:
		[\mu_1^2+\nu_1^2+\mu_2^2+\nu_2^2+\mu_3^2+\nu_3^2+\mu_4^2+\nu_4^2=1
		\wedge \tfrac{337}{625}\nu_1^4 \ + \\
	& \qquad \tfrac{3318403704685565836307974101662}{5684341886080801486968994140625}\nu_1^2\nu_2^2
		+\tfrac{5017502987841674535674567823313}{5684341886080801486968994140625}\nu_2^4
		+\tfrac{14}{25}\nu_1^2\nu_3^2 \ + \\
	& \qquad \tfrac{10033612198548867359598636674}{9094947017729282379150390625}\nu_2^2\nu_3^2 
		+\nu_3^4 +\tfrac{674}{625}\nu_1^2\nu_4^2
		+\tfrac{5494274924825481229075961726}{9094947017729282379150390625}\nu_2^2\nu_4^2 \ + \\
	& \qquad \tfrac{14}{25}\nu_3^2\nu_4^2+\nu_4^4+\tfrac{674}{625}\nu_1^2\mu_1^2
		+\tfrac{3318403704685565836307974101662}{5684341886080801486968994140625}\nu_2^2\mu_1^2
		+\tfrac{14}{25}\nu_3^2\mu_1^2+\tfrac{674}{625}\nu_4^2\mu_1^2 \ + \\
	& \qquad \tfrac{337}{625}\mu_1^4
		+\tfrac{3318403704685565836307974101662}{5684341886080801486968994140625}\nu_1^2\mu_2^2
		+\tfrac{10035005975683349071349135646626}{5684341886080801486968994140625}\nu_2^2\mu_2^2 \ + \\
	& \qquad \tfrac{10033612198548867359598636674}{9094947017729282379150390625}\nu_3^2\mu_2^2
		+\tfrac{5494274924825481229075961726}{9094947017729282379150390625}\nu_4^2\mu_2^2 \ + \\
	& \qquad \tfrac{3318403704685565836307974101662}{5684341886080801486968994140625}\mu_1^2\mu_2^2
		+\tfrac{5017502987841674535674567823313}{5684341886080801486968994140625}\mu_2^4
		+\tfrac{14}{25}\nu_1^2\mu_3^2 \ + \\
	& \qquad \tfrac{10033612198548867359598636674}{9094947017729282379150390625}\nu_2^2\mu_3^2
		+2\nu_3^2\mu_3^2+\tfrac{14}{25}\nu_4^2\mu_3^2 +\tfrac{14}{25}\mu_1^2\mu_3^2 \ + \\
	& \qquad \tfrac{10033612198548867359598636674}{9094947017729282379150390625}\mu_2^2\mu_3^2
		+\mu_3^4+\tfrac{674}{625}\nu_1^2\mu_4^2
		+\tfrac{5494274924825481229075961726}{9094947017729282379150390625}\nu_2^2\mu_4^2 \ + \\
	& \qquad \tfrac{14}{25}\nu_3^2\mu_4^2+2\nu_4^2\mu_4^2 +\tfrac{674}{625}\mu_1^2\mu_4^2
		+\tfrac{5494274924825481229075961726}{9094947017729282379150390625}\mu_2^2\mu_4^2
		+\tfrac{14}{25}\mu_3^2\mu_4^2 +\mu_4^4 \le \tau^2].
	\end{aligned}
\]

	By \textsc{Reduce}~\cite{DoS97},
	the fidelity-quantifier formula $\QQ_{\le 3351/5000}(\phi_4)$
	is decided to be true
	while $\QQ_{\le 67/100}(\phi_4)$ is false.
	In other words,
	$\mfid(\Delta(\phi_4))$ is in $(\tfrac{67}{100},\tfrac{3351}{5000}]$,
	which entails that
	at least $67\%$ of the original MAC and proper IP addresses at $s_3$
	would be delivered at the terminal $s_4$ or $s_5$ within 15 steps
	through the noisy channel $\QMC_1$.
	Besides, by~\cite{FYY13}, we can compute that
	it has probability at least $\tfrac{53}{100}$ to reach $s_4$ or $s_5$ within 15 steps,
	whose square-root can be proven to be an upper bound of the fidelity,
	i.e. $(\tfrac{53}{100})^{1/2} > \tfrac{3351}{5000}$, no lower bound.
	So it is less precise than ours
	in characterizing the similarity degree of the two MAC addresses.
	
	For the formulas $\phi_1$ and $\phi_2$, we have that
	both $\QQ_{=0}(\phi_1)$ and $\QQ_{=0}(\phi_2)$ hold,
	since there are some pure states,
	whose support falls into the BSCC subspaces w.r.t. $\FFF_{\neg\ok}$.
	For instance, $\op{s_3}{s_3}\otimes\op{1,1}{1,1}$
	is tansformed to $\op{s_4}{s_4}\otimes Q(s_3,s_4)(\op{1,1}{1,1})
	=\tfrac{337}{625}\op{s_4}{s_4}\otimes\op{1,1}{1,1}$,
	whose support itself forms a BSCC subspace;
	and to $\op{s_0}{s_0}\otimes Q(s_3,s_0)(\op{1,1}{1,1})
	=\tfrac{288}{625}\op{s_0}{s_0}\otimes\op{1,1}{1,1}$,
	whose support falls into the BSCC subspace
	$\spn(\ket{s_0}\otimes\ket{1,1},\ket{s_1}\otimes\ket{1,+})$.
	Besides, we have that
	both $\QQ_{>3351/5000}(\phi_3)$ and $\QQ_{\le 6703/10000}(\phi_3)$ hold,
	as the bounded-until formula approaches the unbounded-until one.
\end{example}

\begin{remark}
	When the initial density operator $\rho$ is given
	and all the entries in the Kraus operators of $Q(s,t)$ with $s,t\in S$ are rational,
	it would be in polynomial time to decide $(s,\rho)\models\QQ_{\sim\tau}(\phi)$,
	since the time-consuming quantifier elimination is saved then.
	It is consistent with the existing work~\cite{YYF13}.
\end{remark}

\paragraph{Implementation}
	We have implemented the presented method on the platform \textsc{Mathematica},
	incorported with the package \textsc{Reduce}~\cite{DoS97}.
	Using caching mechanism,
	we divide the whole computation procedure into two subprocedures:
	\begin{enumerate}
		\item the synthesizing subprocedure to prepare information
		about the given QMC and QCTL formula,
		including quantum state information,
		matrix representations of transition super-operators,
		and removal of the direct-sum of all BSCC subspaces; and
		\item the deciding subprocedure for fidelity-quantifier formulas.
	\end{enumerate}
	Thus we can for instance ensure good interactivity in specific situation.
	Under a PC with Intel Core i7-6700 CPU and 8 GB RAM,
	the overall performance of our running examples is acceptable:
	the total time consumption is within a few seconds;
	the synthesizing subprocedure consumes nearly $30$ MB memory,
	and besides that,
	the deciding subprocedure consumes at most $50$ MB.
	%Also, 
	%some optimization techniques could be further taken into account
	%to reduce complexity
	%when we deal with matrix power and synthesis of super-operators on a particular path. 
	Finally, we have to address that
	the fidelity computation for
	the QMC with a \emph{concrete} initial classical--quantum state
	is always much efficient (usually within 1 second);
	while the fidelity computation for
	the QMC with a \emph{parametric} initial classical--quantum state may be inefficient,
	since in the worst case the quantifier elimination is exponential time.
	Detailed calculation procedure is available at
	\texttt{https://github.com/melonysuga/PaperFidelityExamples-.git}.

%\begin{remark}
%	In fact, the optimal quantum error-correction (QEC) problem
%	that for a given error super-operator $\EEE$,
%	what is the optimal recovery super-operator $\RRR$
%	that maximizes the (minimum) fidelity over all initial quantum states $\rho$,
%	i.e. $\mfid(\RRR \circ \EEE)=\min_\rho \fid(\RRR \circ \EEE,\rho)$
%	is harder than the fidelity computation problem in this paper.
%	Yamamoto~\textit{et~al.} applied the techniques of
%	semi-definite programming and sum of squares
%	to the QEC problem in~\cite{YHT05,Yam09}.
%	These existing methods are insufficient to
%	answer the decision problem $s\models\Phi$
%	for the quantum system of $2$-qubit or more as we expect,
%	due to the numerical errors caused by approximation
%	or the theoretical bottleneck of SOS.
%	Compared with them, however,
%	our method adopts the technique of quantifier elimination
%	effective for quantum systems of arbitrarily many qubits.
%\end{remark}

\section{Conclusion}\label{S7}
In this paper,
we introduced a quantum analogy of computation tree logic (QCTL),
which consists of state formulas and path formulas.
A model checking algorithm was presented over the quantum Markov chains (QMCs).
We gave a simple polynomial time procedure
that could remove all fixed-points w.r.t. a super-operator.
We then synthesized the super-operators of path formulas
using explicit matrix representations,
and decided the fidelity-quantifier formulas
by a reduction to quantifier elimination in the existential theory of the reals.
Finally, the QCTL formulas were shown to be decidable in exponential time.

We believe that the proposed method could be extended to:
\begin{itemize}
	\item synthesize the SOVM for the general multiphase until formula
	$\Phi_1 \ntl^{\mathbb{I}_1} \Phi_2 \ntl^{\mathbb{I}_2} \cdots
	\ntl^{\mathbb{I}_{k-1}} \Phi_k$
	with proper time intervals $\mathbb{I}_i$ as in~\cite{XZJ+16};
	\item synthesize the SOVM for the conjunction $\phi_1 \wedge \phi_2$,
	so that the conditional fidelity,
	similar to conditional probability~\cite{AvR08,GXZ+13},
	could be established;
	\item synthesize the SOVM for the negation $\neg\phi$,
	so that the safety property
	$\Box\,\Phi = \neg (\true\,\ntl \neg\Phi)$
	could be analyzed; and
	\item decide the analogy of SOVM-quantifier formula over parametric QMCs.
	The positive-operator valued measure (POVM) would be a key tool to attack it.
\end{itemize}



\bibliographystyle{elsarticle-harv} 
\bibliography{fidelity}

\begin{thebibliography}{31}
\expandafter\ifx\csname natexlab\endcsname\relax\def\natexlab#1{#1}\fi
\expandafter\ifx\csname url\endcsname\relax
  \def\url#1{\texttt{#1}}\fi
\expandafter\ifx\csname urlprefix\endcsname\relax\def\urlprefix{URL }\fi

\bibitem[{Andova et~al.(2004)Andova, Hermanns, and Katoen}]{AHK04}
Andova, S., Hermanns, H., Katoen, J.-P., 2004. Discrete-time rewards
  model-checked. In: Larsen, K.~G., Niebert, P. (Eds.), Formal Modeling and
  Analysis of Timed Systems: First International Workshop, FORMATS 2003. Vol.
  2791 of LNCS. Springer, pp. 88--104.

\bibitem[{Andr{\'e}s and van Rossum(2008)}]{AvR08}
Andr{\'e}s, M.~E., van Rossum, P., 2008. Conditional probabilities over
  probabilistic and nondeterministic systems. In: Ramakrishnan, C.~R., Rehof,
  J. (Eds.), Tools and Algorithms for the Construction and Analysis of Systems:
  14th International Conference, TACAS 2008. Vol. 4963 of LNCS. Springer, pp.
  157--172.

\bibitem[{Arute et~al.(2019)Arute, Arya, Babbush, Bacon, Bardin, Barends,
  Biswas, Boixo, Brandao, Buell, Burkett, Chen, Chen, Chiaro, Collins,
  Courtney, Dunsworth, Farhi, Foxen, Fowler, Gidney, Giustina, Graff, Guerin,
  Habegger, Harrigan, Hartmann, Ho, Hoffmann, Huang, Humble, Isakov, Jeffrey,
  Jiang, Kafri, Kechedzhi, Kelly, Klimov, Knysh, Korotkov, Kostritsa, Landhuis,
  Lindmark, Lucero, Lyakh, Mandr\'{a}, McClean, McEwen, Megrant, Mi,
  Michielsen, Mohseni, Mutus, Naaman, Neeley, Neill, Niu, Ostby, Petukhov,
  Platt, Quintana, Rieffel, Roushan, Rubin, Sank, Satzinger, Smelyanskiy, Sung,
  Trevithick, Vainsencher, Villalonga, White, Yao, Yeh, Zalcman, Neven, and
  Martinis}]{AAB+19}
Arute, F., Arya, K., Babbush, R., Bacon, D., Bardin, J.~C., Barends, R.,
  Biswas, R., Boixo, S., Brandao, F. G. S.~L., Buell, D.~A., Burkett, B., Chen,
  Y., Chen, Z., Chiaro, B., Collins, R., Courtney, W., Dunsworth, A., Farhi,
  E., Foxen, B., Fowler, A., Gidney, C., Giustina, M., Graff, R., Guerin, K.,
  Habegger, S., Harrigan, M.~P., Hartmann, M.~J., Ho, A., Hoffmann, M., Huang,
  T., Humble, T.~S., Isakov, S.~V., Jeffrey, E., Jiang, Z., Kafri, D.,
  Kechedzhi, K., Kelly, J., Klimov, P.~V., Knysh, S., Korotkov, A., Kostritsa,
  F., Landhuis, D., Lindmark, M., Lucero, E., Lyakh, D., Mandr\'{a}, S.,
  McClean, J.~R., McEwen, M., Megrant, A., Mi, X., Michielsen, K., Mohseni, M.,
  Mutus, J., Naaman, O., Neeley, M., Neill, C., Niu, M.~Y., Ostby, E.,
  Petukhov, A., Platt, J.~C., Quintana, C., Rieffel, E.~G., Roushan, P., Rubin,
  N.~C., Sank, D., Satzinger, K.~J., Smelyanskiy, V., Sung, K.~J., Trevithick,
  M.~D., Vainsencher, A., Villalonga, B., White, T., Yao, Z.~J., Yeh, P.,
  Zalcman, A., Neven, H., Martinis, J.~M., 2019. Quantum supremacy using a
  programmable superconducting processor. Nature 574, 505--510.

\bibitem[{Baier and Katoen(2008)}]{BaK08}
Baier, C., Katoen, J.-P., 2008. Principles of Model Checking. MIT Press.

\bibitem[{Basu et~al.(2006)Basu, Pollack, and Roy}]{BPR06}
Basu, S., Pollack, R., Roy, M.-F., 2006. Algorithms in Real Algebraic Geometry,
  2nd Edition. Springer.

\bibitem[{Bennett and Brassard(1984)}]{BeB84}
Bennett, C.~H., Brassard, G., 1984. Quantum cryptography: Public key
  distribution and coin tossing. In: Proc. of IEEE International Conference on
  Computers, Systems and Signal Processing, 1984. IEEE Computer Society, pp.
  175--179.

\bibitem[{Clarke et~al.(1999)Clarke, Grumberg, and Peled}]{CGP99}
Clarke, E.~M., Grumberg, O., Peled, D.~A., 1999. Model Checking. MIT Press.

\bibitem[{de~Moura and Bj{\o}rner(2008)}]{dMB08}
de~Moura, L., Bj{\o}rner, N., 2008. \textsc{Z3}: An efficient {SMT} solver. In:
  Ramakrishnan, C.~R., Rehof, J. (Eds.), Tools and Algorithms for the
  Construction and Analysis of Systems: 14th International Conference, TACAS
  2008. Vol. 4963 of LNCS. Springer, pp. 337--340.

\bibitem[{Dehnert et~al.(2017)Dehnert, Junges, Katoen, and Volk}]{DJK+17}
Dehnert, C., Junges, S., Katoen, J.-P., Volk, M., 2017. A \textsc{Storm} is
  coming: A modern probabilistic model checker. In: Majumdar, R., Kuncak, V.
  (Eds.), Computer Aided Verification: 29th International Conference, {CAV}
  2017, Part {II}. Vol. 10427 of LNCS. Springer, pp. 592--600.

\bibitem[{Dolzmann and Sturm(1997)}]{DoS97}
Dolzmann, A., Sturm, T., 1997. \textsc{Redlog}: Computer algebra meets computer
  logic. ACM SIGSAM Bulletin 31~(2), 2--9.

\bibitem[{Feng et~al.(2017)Feng, Hahn, Turrini, and Ying}]{FHT+17}
Feng, Y., Hahn, E.~M., Turrini, A., Ying, S., 2017. Model checking
  $\omega$-regular properties for quantum {Markov} chains. In: Meyer, R.,
  Nestmann, U. (Eds.), 28th International Conference on Concurrency Theory,
  {CONCUR} 2017. Vol.~85 of LIPIcs. Schloss Dagstuhl --- Leibniz-Zentrum
  f{\"{u}}r Informatik, pp. 35:1--35:16.

\bibitem[{Feng et~al.(2013)Feng, Yu, and Ying}]{FYY13}
Feng, Y., Yu, N., Ying, M., 2013. Model checking quantum {Markov} chains.
  Journal of Computer and System Sciences 79~(7), 1181--1198.

\bibitem[{Gao et~al.(2013)Gao, Xu, Zhan, and Zhang}]{GXZ+13}
Gao, Y., Xu, M., Zhan, N., Zhang, L., 2013. Model checking conditional {CSL}
  for continuous-time {Markov} chains. Information Processing Letters
  113~(1-2), 44--50.

\bibitem[{Gay et~al.(2006)Gay, Nagarajan, and Papanikolaou}]{GNP06}
Gay, S.~J., Nagarajan, R., Papanikolaou, N., 2006. Probabilistic model-checking
  of quantum protocols. In: Proc. 2nd International Workshop on Developments in
  Computational Models.

\bibitem[{Gay et~al.(2008)Gay, Nagarajan, and Papanikolaou}]{GNP08}
Gay, S.~J., Nagarajan, R., Papanikolaou, N., 2008. \textsc{QMC}: A model
  checker for quantum systems. In: Gupta, A., Malik, S. (Eds.), Computer Aided
  Verification, 20th International Conference, {CAV} 2008. Vol. 5123 of LNCS.
  Springer, pp. 543--547.

\bibitem[{Grover(1996)}]{Gro96}
Grover, L.~K., 1996. A fast quantum mechanical algorithm for database search.
  In: Proc. 28th Annual {ACM} Symposium on the Theory of Computing. {ACM}, pp.
  212--219.

\bibitem[{Guan et~al.(2018)Guan, Feng, and Ying}]{GFY18}
Guan, J., Feng, Y., Ying, M., 2018. Decomposition of quantum {Markov} chains
  and its applications. Journal of Computer and System Sciences 95, 55--68.

\bibitem[{Hahn et~al.(2014)Hahn, Li, Schewe, Turrini, and Zhang}]{HLS+14}
Hahn, E.~M., Li, Y., Schewe, S., Turrini, A., Zhang, L., 2014.
  \textsc{iscasMc}: A web-based probabilistic model checker. In: Jones, C.,
  Pihlajasaari, P., Sun, J. (Eds.), FM 2014: Formal Methods---19th
  International Symposium. Vol. 8442 of LNCS. Springer, pp. 312--317.

\bibitem[{Han et~al.(2009)Han, Katoen, and Damman}]{HKD09}
Han, T., Katoen, J.-P., Damman, B., 2009. Counterexample generation in
  probabilistic model checking. IEEE Transactions on Software Engineering
  35~(2), 241--257.

\bibitem[{Hansson and Jonsson(1989)}]{HaJ89}
Hansson, H., Jonsson, B., 1989. A framework for reasoning about time and
  reliability. In: Proc. IEEE Real-Time Systems Symposium, 1989. IEEE Computer
  Society, pp. 102--111.

\bibitem[{Harrow et~al.(2009)Harrow, Hassidim, and Lloyd}]{HHL09}
Harrow, A.~W., Hassidim, A., Lloyd, S., 2009. Quantum algorithm for solving
  linear systems of equations. Physical Review Letters 103~(15), article no.
  150502.

\bibitem[{Istr\v{a}\c{t}escu(2001)}]{Ist01}
Istr\v{a}\c{t}escu, V.~I., 2001. Fixed Point Theory: An Introduction. Springer.

\bibitem[{Kwiatkowska et~al.(2011)Kwiatkowska, Norman, and Parker}]{KNP11}
Kwiatkowska, M., Norman, G., Parker, D., 2011. \textsc{PRISM} 4.0: Verification
  of probabilistic real-time systems. In: Gopalakrishnan, G., Qadeer, S.
  (Eds.), Computer Aided Verification: 23rd International Conference, CAV 2011.
  Vol. 6806 of LNCS. Springer, pp. 585--591.

\bibitem[{Li and Feng(2015)}]{LiF15}
Li, L., Feng, Y., 2015. Quantum {Markov} chains: Description of hybrid systems,
  decidability of equivalence, and model checking linear-time properties.
  Information and Computation 244, 229--244.

\bibitem[{Nielsen and Chuang(2000)}]{NiC00}
Nielsen, M.~A., Chuang, I.~L., 2000. Quantum Computation and Quantum
  Information. Cambridge University Press.

\bibitem[{Shor(1994)}]{Sho94}
Shor, P.~W., 1994. Algorithms for quantum computation: Discrete logarithms and
  factoring. In: Proc. 35th Annual Symposium on Foundations of Computer
  Science. {IEEE} Computer Society, pp. 124--134.

\bibitem[{Tarski(1951)}]{Tar51}
Tarski, A., 1951. A Decision Method for Elementary Algebra and Geometry, 2nd
  Edition. University of California Press.

\bibitem[{Wootters and Zurek(1982)}]{WoZ82}
Wootters, W.~K., Zurek, W.~H., 1982. A single quantum cannot be cloned. Nature
  299, 802--803.

\bibitem[{Xu et~al.(2016)Xu, Zhang, Jansen, Zhu, and Yang}]{XZJ+16}
Xu, M., Zhang, L., Jansen, D.~N., Zhu, H., Yang, Z., 2016. Multiphase until
  formulas over {Markov} reward models: An algebraic approach. Theoretical
  Computer Science 611, 116--135.

\bibitem[{Ying et~al.(2013{\natexlab{a}})Ying, Yu, Feng, and Duan}]{YYF13}
Ying, M., Yu, N., Feng, Y., Duan, R., 2013{\natexlab{a}}. Verification of
  quantum programs. Science of Computer Programming 78~(9), 1679--1700.

\bibitem[{Ying et~al.(2013{\natexlab{b}})Ying, Feng, Yu, and Ying}]{YFY+13}
Ying, S., Feng, Y., Yu, N., Ying, M., 2013{\natexlab{b}}. Reachability
  probabilities of quantum {Markov} chains. In: D'Argenio, P.~R., Melgratti,
  H.~C. (Eds.), {CONCUR} 2013: Concurrency Theory---24th International
  Conference. Vol. 8052 of LNCS. Springer, pp. 334--348.

\end{thebibliography}




%%\bibitem{}

%%\end{thebibliography}
\end{document}